\documentclass[%
aip,
amsmath,amssymb,
reprint,%
]{revtex4-1}

\usepackage{graphicx}
\usepackage{dcolumn}
\usepackage{bm}

\usepackage[utf8]{inputenc}
\usepackage[T1]{fontenc}
\usepackage{mathptmx}

\usepackage[colorlinks=false,linkcolor=black]{hyperref}
\usepackage{xcolor, soul} 
\sethlcolor{green} 

\begin{document}

	\title{Realization of all logic gates and memory latch in the SC-CNN cell of the simple nonlinear MLC circuit}
	
	\author{P. Ashokkumar} 
	\email{ak3phys@gmail.com}
	\affiliation{PG \& Research Department of Physics, Nehru Memorial College (Autonomous), Affiliated to Bharathidasan University, Puthanampatti, Tiruchirappalli - 621 007, India.}
	
	\author{M. Sathish Aravindh} 
	\email{sathisharavindhm@gmail.com}
	\affiliation{PG \& Research Department of Physics, Nehru Memorial College (Autonomous), Affiliated to Bharathidasan University, Puthanampatti, Tiruchirappalli - 621 007, India.}
	\affiliation{Department of Nonlinear Dynamics, School of Physics, Bharathidasan University, Tiruchirappalli - 620 024, India.}
	
	\author{A. Venkatesan}
	\email{av.phys@gmail.com}
	\affiliation{PG \& Research Department of Physics, Nehru Memorial College (Autonomous), Affiliated to Bharathidasan University, Puthanampatti, Tiruchirappalli - 621 007, India.}

	\author{M. Lakshmanan}
	\email{lakshman.cnld@gmail.com}
	\affiliation{Department of Nonlinear Dynamics, School of Physics, Bharathidasan University, Tiruchirappalli - 620 024, India.}
		
	\begin{abstract}
	We investigate the State-Controlled Cellular Neural Network (SC-CNN) framework of Murali-Lakshmanan-Chua (MLC) circuit system subjected to two logical signals. By exploiting the attractors generated by this circuit in different regions of phase-space, we show that the nonlinear circuit is capable of producing all the logic gates, namely OR, AND, NOR, NAND, Ex-OR and Ex-NOR gates available in digital systems. Further the circuit system emulates three-input gates and Set-Reset flip-flop logic as well. Moreover, all these logical elements and flip-flop are found to be tolerant to noise. These phenomena are also experimentally demonstrated. Thus our investigation to realize all logic gates and memory latch in a nonlinear circuit system paves the way to replace or complement the existing technology with a limited number of hardware.
	
	\end{abstract}
	\maketitle

	\begin{quotation}
		Exploiting the hopping of trajectories in different wells of nonlinear systems, it is found that these  systems possess the ability to produce all kinds of logic operations. Specifically  in the present work, we show that if one uses two square waves in an aperiodic  manner as input to the SC-CNN based simple nonlinear Murali-Lakshmanan-Chua(MLC) circuit, the response produces all the logic gates and RS flip-flop which are also experimentally demonstrated. Our investigation to realize all the logic gates and memory latch in a nonlinear circuit system paves the way to replace or complement the existing technology with a limited number of hardware components. 
	\end{quotation}

	\section{Introduction}
	\label{sec1}

Logic gates are building blocks of any digital circuit and computer architecture. In general, following Boolean algebra, the logic operations are performed by converting given two inputs into a single logical output. For any logic operation, both the inputs and outputs have two states, namely 'ON' or 'TRUE' and 'OFF' or 'FALSE' states \cite{mano2003computer}. The reliability of logic operations depends on the reliable operations of systems chosen. Since the current demand for miniaturization of logic devices, speed of computation and low-power consumption devices, it is inevitable to design an appropriate system which is able to produce noise-immune gates. As a result, for the past two decades or more, several schemes are being proposed such as DNA/RNA computing \cite{adleman1994molecular,collier1999electronically}, quantum computing \cite{ladd2010quantum,nielsen2002quantum}, nano computing \cite{bachtold2001logic} and nonlinear dynamics based computing in order to replace or complement the existing computer architecture based on silicon chips\cite{hopfield1982neural,sinha1998dynamics, sinha1999computing, prusha1999nonlinearity, murali2009realization, murali2009reliable, sinha2009exploiting, guerra2010noise, worschech2010universal, zamora2010numerical, zhang2010effect, bulsara2010logical, singh2011enhancement, dari2011creating, dari2011noise, storni2012manipulating, roychowdhury2015boolean, kohar2017implementing, venkatesh2017implementation, venkatesh2017design, neves2017noise, kia2017nonlinear,murali2018chaotic,murali2018coupling, aravindh2018strange, sathish2020realisation}. Among these methods, nonlinear dynamics based computing can make reliable and re-configurable computer architecture because the underlying nonlinear systems posses a large number of basic functions.  

Utilizing the flexibility of nonlinear dynamical systems for storing, communicating and processing of information in computer architecture has been an active area of research in nonlinear dynamics. In this connection, Hopfield had constructed a memory device using artificial neural network to store and retrieve information \cite{hopfield1982neural}. Sinha and Ditto proposed a chaos-computing scheme to emulate different logic elements\cite{sinha1998dynamics, sinha1999computing}. In an optimal window of noise, it was observed the possibility of occurrence of logic behavior and this phenomenon is termed as logical stochastic resonance(LSR)\cite{murali2009realization, murali2009reliable, sinha2009exploiting,bulsara2010logical}. The present authors employed strange nonchaotic attractors to build dynamical logic gates \cite{aravindh2018strange,sathish2020realisation,sathish2020route}. Besides, self-sustained oscillators can function as latches and registers if Boolean logic states are associated with the phases of the oscillator signals\cite{roychowdhury2015boolean}. Heteroclinic computing is another nonlinear phenomenon based computation using a collective system of nonlinear oscillators \cite{ashwin2004encoding, ashwin2005discrete}.        

	Not only nonlinear dynamics based computing complements existing silicon based technology, there also exist several efforts to extend computation techniques to other domains such as optical\cite{singh2011enhancement}, chemical\cite{sinha2009exploiting, de2009implementation,stevens2012time}, physical\cite{aravindh2018strange}, mechanical\cite{mahboob2011interconnect}, biological\cite{gerstung2009noisy,ando2011synthetic}, molecular\cite{collier1999electronically,kompa2001molecular} and other areas of science\cite{motoike2001real,sinha2002flexible,norrell2009boolean,zhang2010effect,mozeika2010noisy,zhang2012logical,miyamoto2010resonant}. Instead of needing multiple hardware for different types of computations, nonlinear dynamical systems can act as processors of a flexibly configured and reconfigured device to produce different logic gates \cite{worschech2010universal}. In practice, the generation of nonidealitic and ambient noise restricts the ability to obtain different logic gates in these systems \cite{murali2009reliable}. Thus it is essential to choose appropriate nonlinear-dynamics based computing systems to overcome these odds.
	
	Further, most of the previous studies have focused to produce OR(NOR) and AND(NAND) logic gates \cite{sinha1998dynamics, sinha1999computing, prusha1999nonlinearity, murali2009realization, murali2009reliable, kohar2012noise, kohar2014enhanced, kohar2017implementing, venkatesh2017implementation, venkatesh2017design, aravindh2018strange, sathish2020realisation}. Obtaining Ex-OR (simply XOR) and Ex-NOR (XNOR) in dynamical systems are equally important since these gates are the basis of ubiquitous bit-by-bit addition \cite{storni2012manipulating}. A half adder consists of an XOR gate and AND gate. Other uses of XOR gate include subtractor, comparator and controlled inverter. XOR and XNOR gates are usually obtained by the concatenation of NOR and NAND gates. 
	
	\textcolor{black}{In the literature, it has been shown that in several nonlinear-dynamics based concepts like chaos computing \cite{ditto2008chaos}, heteroclinic network for computation \cite{ashwin2004encoding, ashwin2005discrete, bick2009dynamical, neves2012computation, neves2017noise}, etc. the richness of nonlinear dynamics can be exploited to obtain flexible and reconfigurable logic gates including XOR gates. In this connection, Sinha \emph{et al.} have demonstrated the flexible parallel implementation of logic gates using chaotic elements \cite{sinha2002flexible}. Peng \emph{et al.} have explored piecewise-linear systems to construct all dynamical logic gates \cite{peng2008harnessing, peng2010dynamic}. Campos-Canton \emph{et al.} have reported electronics experiments to obtain NOR, NAND and XOR gates in a piece-wise linear system \cite{campos2010simple}. In ref.\cite{cafagna2006chaos,campos2012set} the equation of the plane in analytical geometry  has been used to build a contribution of SR flip-flop and basic logic gates \cite{cafagna2006chaos, campos2012set}.  Storni  \emph{et al.} have investigated LSR by extending the analysis to a three well potential to realize XOR logic gates \cite{storni2012manipulating}. In ref.\cite{canton2017method}, the authors have proposed a method and circuit for integrating a programmable matrix in the field of reconfigurable logic gates employing a nonlinear system and an efficient programmable rewiring \cite{canton2017method}. Also Campos-Canton \emph{et al.} have reported a parameterized method to design multivibrator circuit via Chua's circuit system \cite{campos2012multivibrator}.} 
	
	\textcolor{black}{Further, Murali \emph{et al.} have shown that if one applies two low amplitude square waves as inputs to a two-state system, the response of the system produces a logical output (NOR/OR) with a probability controlled by the interplay between noise and bistable dynamics of the system. That is the interplay of nonlinearity and noise produces a flexible and realizable logic behavior. The authors have termed this phenomenon as Logical Stochastic Resonance (LSR) \cite{bulsara2010logical}. For the past few years LSR has been realized in many nonlinear systems such as a nanoscale device \cite{guerra2010noise}, resonant tunnel diodes \cite{worschech2010universal}, a vertical cavity surface emitting laser \cite{zamora2010numerical, zhang2010effect}, a polarization bistable laser \cite{singh2011enhancement}, a chemical system \cite{bulsara2010logical}, synthetic gene networks \cite{dari2011creating}, and so on. Recently, two of the present authors  along with Venkatesh employed coupled dynamical systems to build dynamical logic gates by altering the value of the logic inputs \cite{venkatesh2017design, venkatesh2017implementation}.} 
	
	\textcolor{black}{Also Gupta \emph{et al.} have examined the possibility of noise free LSR by driving a two-stable system with periodic forcing instead of random noise \cite{gupta2011noise}. Kohar \emph{et al.} have found that periodic forcing enhances the LSR in noisy bistable systems with periodic forcing instead of random noise \cite{kohar2014enhanced}. By replacing noise with high-frequency harmonics, Venkatesh \emph{et al.}  realized logic operations AND, OR and RS flip-flop in the MLC circuit and they termed this phenomenon as Logical Vibrational Resonance (LVR). LVR has been realized in a two-potential system \cite{venkatesh2016vibrational, venkatesh2017design, venkatesh2017implementation}. Besides logic gates, LSR and LVR can be used to realize SR flip-flops \cite{gui2020enhanced, gui2020set, vincent2021vibrational, murali2021construction}. Then the question arises whether LSR and LVR can produce all the logic gates including XOR gate when one employs a bistable system. One finds that practically this is not possible. The reason for this shortcoming is the bistable nonlinearity of the system. 
	For this case all the logic gates are obtained by assuming that for the input state $ (0,0) $, the response of the bistable system resides in the left well and  for the input state $ (1,1) $ it is in the right well and for the other input state $ (0,1)/(1,0) $ it may be any one of the wells depending upon the logic operations chosen. With these assumptions, bistable nonlinear systems are well suited for producing OR/NOR and AND/NAND logic gates. However these assumptions are not sufficient enough to obtain the other logic gates, namely the XOR and XNOR gates. For example, an XOR gate admits a high logic output only if the inputs are at different logic levels [(0,1) or (1,0)] and low logic output only if the inputs are at the same logic level [(0,0) or (1,1)]. For this logic gate case, the assumption for obtaining logic gates in bistable nonlinear systems leads to a possible loss of information and further the bistable nonlinear systems are unable to hold the condition for XOR/XNOR logic gates.  Thus, the  bistable system is unable to produce XOR/XNOR gates. In this paper, we propose a solution by considering a three-well potential problem so as to implement all the logic elements including XOR gates.  }
	
	It was also shown in several studies that the three well potential nonlinear systems which exhibit logical behaviors are better suited for computational purpose than the \textcolor{black}{logical gates generated by the traditional bistable nonlinear systems when one considers aspects such as noise interference,} waveform smoothness and bit error rate \cite{lu2019decreasing, liu2014new}. Thus it is important to ask the question whether one can realize all the logic gates and memory latches in a single nonlinear circuit system. We address this issue in this paper.
	
	We consider a State-Controlled Cellular Neural Network (SC-CNN) based Murali-Lakshmanan-Chua's (MLC) circuit \cite{gunay2010mlc,gunay2010mlc,swathy2014dynamics}. This circuit is constructed by using two CNN cells and external forces including sinusoidal force, biasing and noise. The existing MLC consists of Chua's diode and a linear resistor, a linear capacitor, and an inductor. The discrete inductor restricts the circuit for fabrications of ICs. Further, CNN is well suited when extending the analysis to the coupled system. Thus SC-CNN MLC is more advantageous than existing MLC in the aspects of hardware realization. We show that this nonautonomous oscillator when subjected to two aperiodic logical signals produces all the logic gates available in digital electronics. We also show that these logic gates are tolerant for an optimal range of noise intensity. Besides, it is reported that the fundamental logical behaviors such as OR/AND/XOR gates can be observed through one of the state variables of the circuit and for the realization of the complimentary logical operations NOR/NAND/XNOR gates the other state variable can be utilized. Further, we report the possibility of the occurrence of both high active SR flip-flop and low SR flip-flop as well because of the parallelism innate in this circuit. One can also further realize three-input gates in the circuit.

	This paper is organized as follows. We discuss the dynamical mechanism for the implementation of logical gates in the three well potential system and present a basic study of the SC-CNN cell of the Murali-Lakshmanan-Chua's (MLC) circuit in Sec.\ref{sec2} and Sec.\ref{sec3}, respectively. We present the experimental realization of CNN cell of the MLC circuit in Sec.\ref{sec4}. We also describe the experimental and numerical results for the realization of OR/NOR and AND/NAND gates, Exclusive-OR (XOR) and Exclusive-NOR (XNOR) logic gates, Set-Reset memory latch, effect of noise and three input gates in Sec.\ref{sec5}. Finally, we conclude our analysis in Sec.\ref{sec6}.

\section{Dynamical mechanism for the implementation of logical gates in three well potential system}
\label{sec2}

Consider a general nonlinear system
\textcolor{black}{ \begin{subequations}
 	\begin{align}
	\ddot{x}& =  -\dfrac{dV(x)}{dx}-h\dot{x}+E+I+\sqrt{D}\xi(t)+fsin(\omega t), ~~ \dot{x}  =  \dfrac{dx}{dt} 	\label{equ1aa} \\
	&\text{Eq.\eqref{equ1aa} can also be rewritten as} \nonumber \\
	\dot{x} &=y, \nonumber \\
	\dot{y}&=-\dfrac{dV(x)}{dx}-h\dot{x}+E+I+\sqrt{D}\xi(t)+fsin z, \nonumber \\
	\dot{z}&=\omega. 	\label{equ1b}	
\end{align}
	\label{equ1a}
\end{subequations}}

	\begin{figure}
	\centering	
	\includegraphics[width=0.8\linewidth]{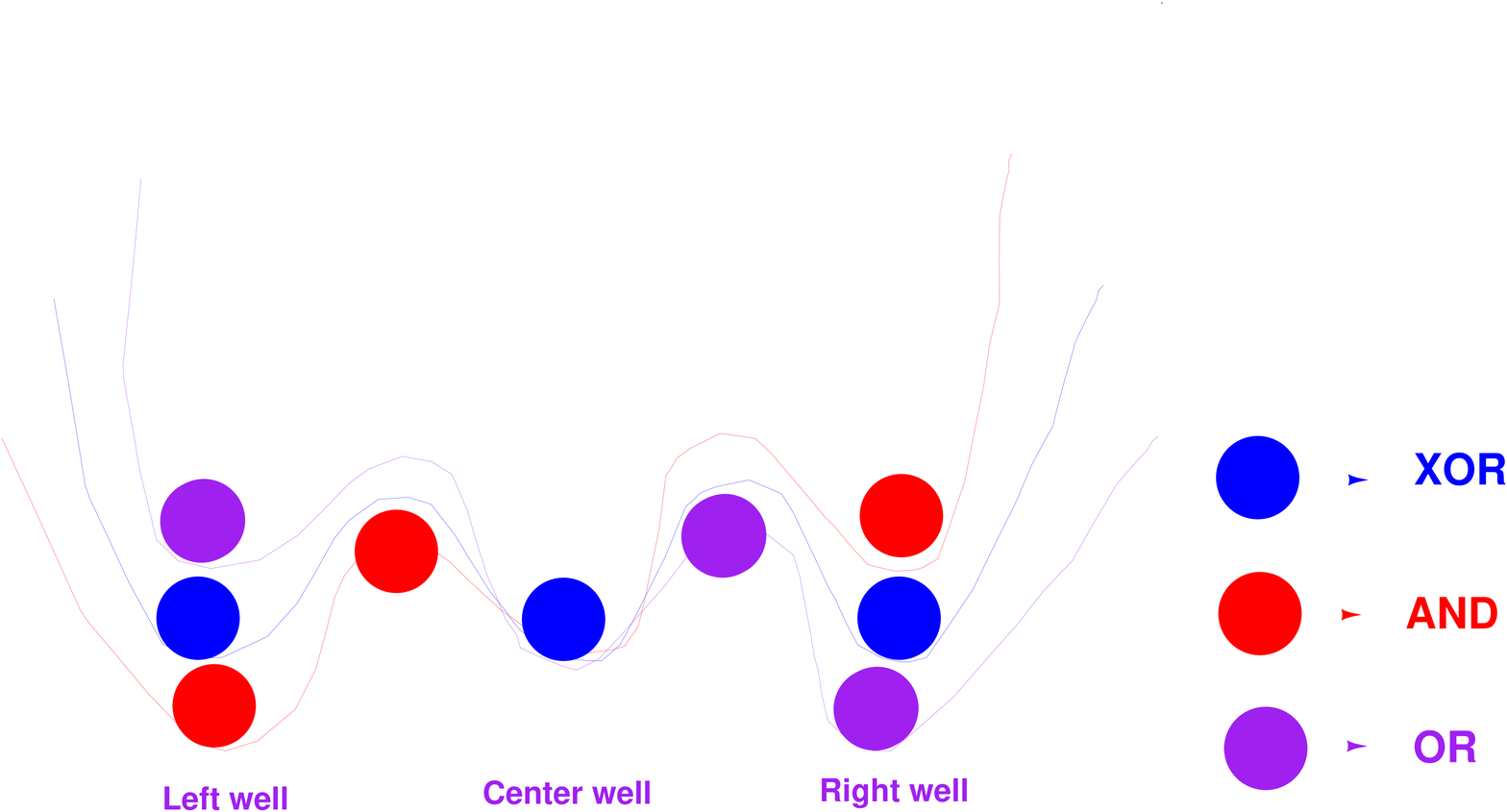} \\
	\caption{Schematic diagram of mechanism for obtaining all the Logic Gates in a triple well system.}
	\label{fig1}
	\end{figure}

Here $ V(x)$ is a potential function of the nonlinear system. Assume that the potential $V(x)$ is three well. It has three potential wells and two potential barriers as shown in Fig.\ref{fig1}. Here  $ E $, $I$, $ \xi(t) $ and $ fsin(\omega t) $ are the bias value, logical input value, Gaussian white noise with intensity `$ D $', and the amplitude of external periodic forcing, respectively. The system \eqref{equ1a} is driven by two square wave signals, namely $ I_{1} $ and $ I_{2} $, which encode the logical inputs and the response of the system \eqref{equ1a} is considered as the logical output. For example, the inputs $ I_{1} $ and $ I_{2} $ take the value $ +\Delta $ for the binary logic number `1' and for $ -\Delta $ the binary logical value is `0'. Then, the four possible combinations viz. (0,0), (0,1)/(1,0) and (1,1) of the input streams $ I_{1}, I_{2} $ are merged into three distinct values $ +2\Delta,0,-2\Delta $ as $ I=I_{1}+I_{2} $. Thus, the resultant input signal $ I $ is a three level aperiodic wave form. Depending on the logic input, the potential function of the system assumes three different forms. The output of logic gates is given by the well in which the state variable `x' resides. More precisely, for OR gate if `x' resides in the left well, we set the logic output as `0', and `1' if it resides in any of the center-well or right-well.  In the XOR gate, the ON output is possible if one and only if one of the inputs to the gate is in the ON state. That is if both the inputs are ON states or both are OFF, an OFF output results. In this circumstance, a bistable system is not able to hold the three input states. Hence when we use a bistable nonlinear system for these kind of cases, it leads to an effective loss of information. To take care of this problem, we propose a solution by generalizing the chosen system from a bistable one to a three well potential system as described above. In particular we extend the scope of implementing the logic gates by increasing the numbers of  potential wells in order to extend the possible input-output associations. For example, for the XOR gate we set the output to be logical '1' if the state value of the system lies between the two local maxima of the potential function, and it is assumed to be  the logical value '0' otherwise. For the XNOR gate we set the output to be logical value '0' if the state of the system resides between the two local maxima of the potential function of the system and it is to be '1' if it resides anywhere in the other two potential wells. The truth table (see Table 1) for different input-output combinations is defined and summarized in Table.\ref{Tab2}.

\begin{table}
	\begin{center}
		\caption{Truth table of logic gates} 
		\vspace{0.2cm}
		\begin{tabular} {|c| c| c| c| c|c|}
			\hline
			\textbf {Logic Gates} & \textbf{(0,0)} & \textbf{(0,1)/(1,0)} & \textbf{(1,1)} \\
			\hline
			\hline
			\textbf{AND} & 0  & 0 &1 \\
			\hline
			\textbf{NAND}  & 1  & 1 & 0  \\
			\hline
			\textbf{OR}  & 0  & 1 &1  \\
			\hline
			\textbf{NOR}  & 1  & 0 &0  \\
			\hline
			\textbf{XOR}  & 0  & 1 &0 \\
			\hline
			\textbf{XNOR}  & 1  & 0 &1  \\
			\hline
			
		\end{tabular}
		\label{Tab1}
	\end{center}
\end{table}

\begin{table}
		\begin{center}
		\caption{Definitions of the outputs for obtaining all logic gates} 
		\vspace{0.2cm}
\begin{tabular} {|c| c| c| c| c|c|}
	\hline
	 \textbf {Logic Gates} & \textbf{Left well} & \textbf{Center well} & \textbf{Right well} \\
	\hline
	\hline
	\textbf{AND}  & OFF  & OFF &ON \\
	\hline
	\textbf{NAND}  & ON  & ON &OFF  \\
	\hline
	\textbf{OR}  & OFF  & ON &ON  \\
	\hline
	\textbf{NOR}  & ON  & OFF &OFF  \\
	\hline
	\textbf{XOR } & OFF  & ON &OFF \\
	\hline
	\textbf{XNOR}  & ON  & OFF & ON  \\
	\hline
	
\end{tabular}
\label{Tab2}
\end{center}
\end{table}
\section{SC-CNN cell of Murali-Lakshmanan-Chua's (MLC) circuit}
\label{sec3}

\begin{figure}
	\centering	
	\includegraphics[width=0.8\linewidth]{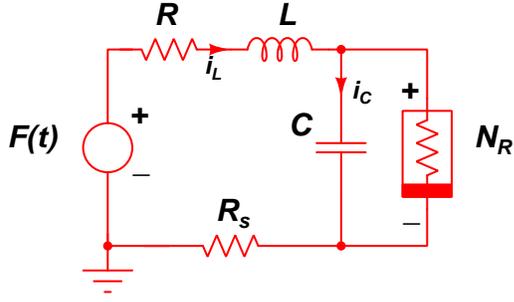} \\
	\caption{Experimental realization of periodically driven Murali-		Lakshmanan-Chua circuit.}
	\label{fig19}
\end{figure}

Cellular Neural Network (CNN), introduced by Chua and Yang  in 1988 \cite{chua1988cellular}, is a n-dimensional array of circuit elements of analog components such as OP-amps, resistors, and capacitors $(n=1,2,3...)$. It is constructed by a large number of intercoupled identical dynamical systems called cells. These cells or nodes are essentially modeled by nonlinear ordinary differential equations. CNN is a relatively simple structure and is easy to be implemented by appropriate electronic circuit. Thus it is a powerful tool for the emulation and implementation of nonlinear dynamical systems having complex dynamics. A generalization of the CNN paradigm in which the CNN cells locally share their outputs as well as their state variables with each other was introduced by Arena \cite{arena1995chua}. This generalization exploits an analog architecture of CNN known as State Controlled-CNN (SC-CNN). Many studies have been reported on designing and implementation of chaotic circuits in terms of SC-CNNs \cite{gunay2010mlc, swathy2014dynamics, luo2016dynamics}. The advantages of a SC-CNN are that they are inductorless and are RC based circuitry only, thereby leading to the realization of hardware and VLSI implementations \cite{manganaro2012cellular}. The other significant applications of these kind of CNNs are parallel computing and low-power consumption. With these motivating facts, in this paper we consider an SC-CNN frame work of the well known MLC circuit. This circuit is essentially perturbed by two aperiodic square waves and as a consequence we show that the resultant output exhibits parallel logic elements and memory latch. In order to discuss this framework further, we first consider the MLC circuit as shown in Fig.\ref{fig19}.

\label{sec2}
\begin{figure}[t]
	\centering
	\includegraphics[width=0.9\linewidth]{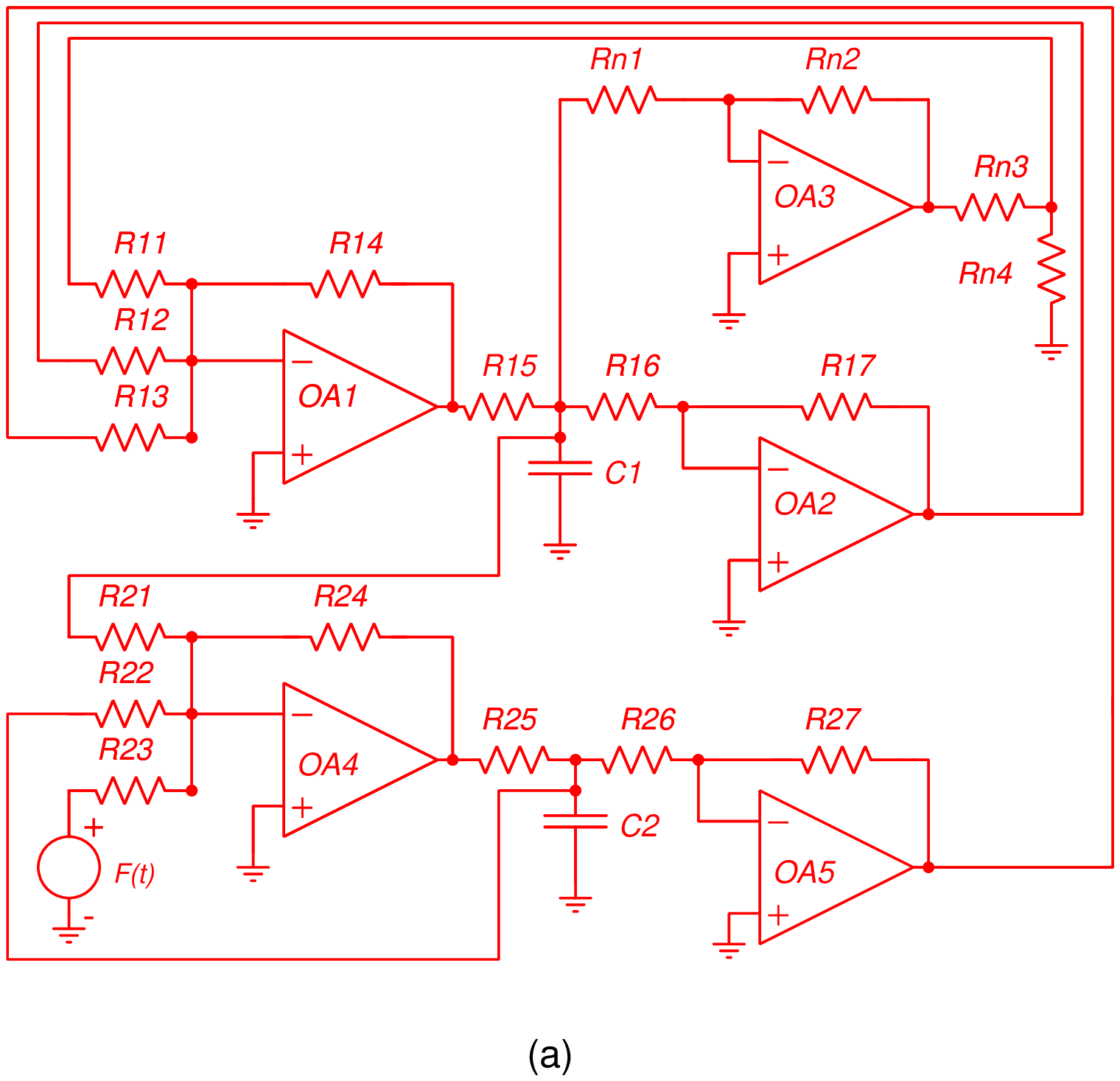}
	\includegraphics[width=0.9\linewidth]{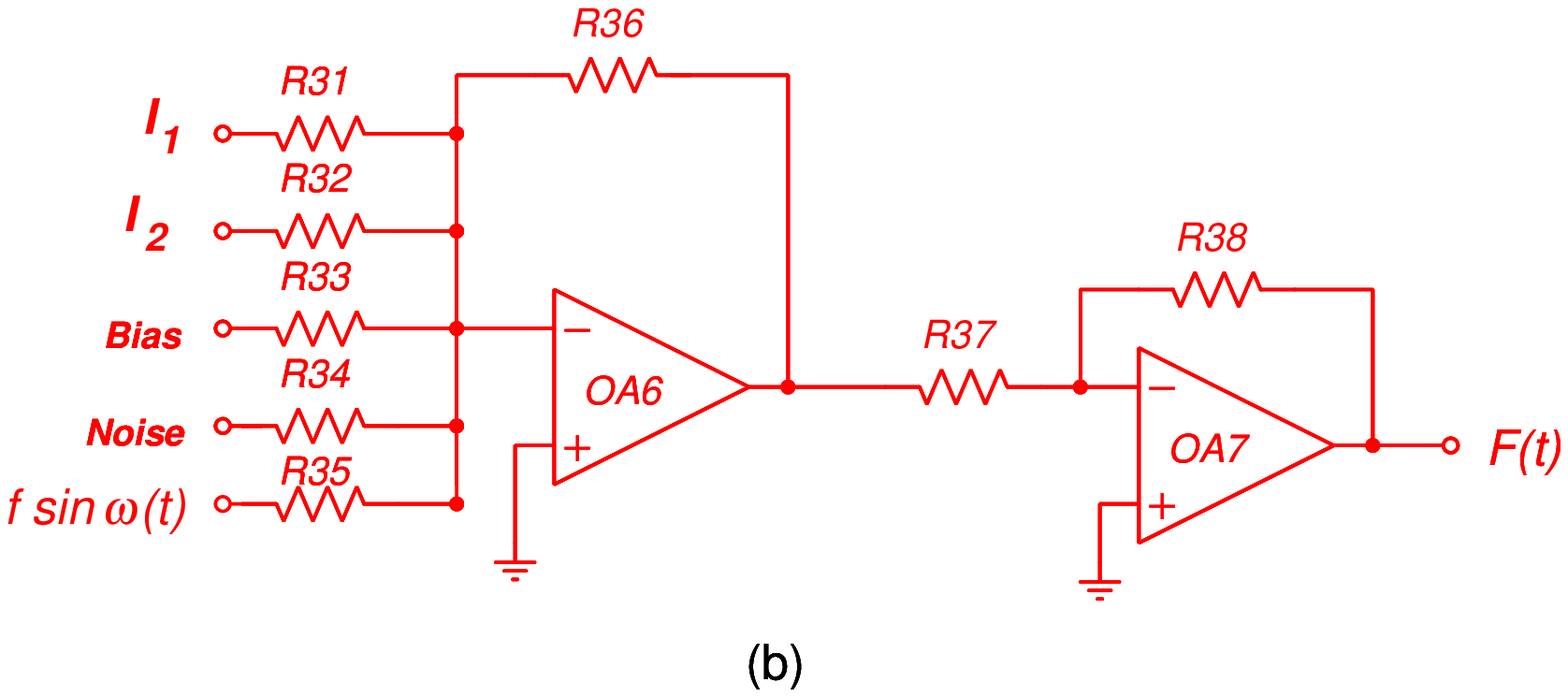}
	\caption{Panel (a) shows the realization of CNN based MLC circuit using five op-amps (OA1-OA5) and 18 Resistors plus two capacitors. The resistors are $ R11-R17, R21-R27,Rn1-Rn4 $. The capacitors are chosen as $ C_{1}=C_{2}=10nF $. Here $ F(t) $ is the input driving signal and panel (b) shows the circuit for generating the driving signal $ F(t) $. Here we use four op-amps (OA6-OA7) and 8 resistors $R31-R38$.}
	\label{fig3}
\end{figure}

It is an established fact in \cite{murali1994simplest, lakshmanan2003chaos} that the normalized form of the well known MLC circuit equation can be written as
\textcolor{black}{
 \begin{subequations}
	\begin{align}
\dot x&= y - h(x), \nonumber \\
\dot y&= -\beta (1+\nu)y - \beta x + E +\sqrt{D}\xi(t) + I + f \sin (\omega t). \label{equ2aa} \\
&\text{or equivalently as} \nonumber \\
\dot x&= y - h(x), \nonumber \\
\dot y&= -\beta (1+\nu)y - \beta x + E +\sqrt{D}\xi(t) + I + f \sin z, \nonumber \\
\dot z&= \omega. \label{equ2b}
\end{align}
\label{equ2}
\end{subequations}}
\noindent Here the overdot indicates time differentiation. \\
\noindent where
\begin{equation}
h(x) =  \left \{
\begin{array} {ll} 
bx + (a - b),   & x > 1,     \\
ax,             & |x| \le 1,     \\
bx - (a - b),   & x < -1.     
\end{array}
\right.
\label{equ3}
\end{equation}

The relationships between the various circuit variables and circuit parameters and the above dynamical variables and parameters can be obtained from Ref.\cite{murali1994simplest, lakshmanan1996chaos, lakshmanan2003chaos}. Earlier studies on the dimensionless version of the circuit when the parameters were fixed at $a=-1.02, \: b=-0.55, \: \gamma = 0.015, \: \beta = 1.0$ and $\omega = 1.0$ were made. The quantity $f$ was varied in these studies \cite{murali1994simplest, lakshmanan2003chaos}. The circuit exhibits various underlying dynamical features including period-doubling route to chaos, intermittent route, strange non-chaotic attractors (SNA), etc \cite{lakshmanan1996chaos,lakshmanan2003chaos}. Three of the present authors have also shown that when the quasiperiodically driven MLC circuit is subjected to two aperiodic and logical square waves, it reproduces logical response in both the SNAs and chaotic regimes of the circuit. The authors have shown that how these attractors are tolerant to noise, so that even one can emulate different logic functions \cite {sathish2020realisation}.   

Following the work of Arena \cite{arena1995chua}, the SC-CNN associated with the MLC circuit can be generalized by the following dimensionless nonlinear state equations, 

\begin{eqnarray}
\dot{x}_{j}&=&-x_{j}+a_j y_{j}+G_{0}+G_{s}+i_{j}, \nonumber \\
\dot{y}_{j}&=&0.5 \times (|x_{j}+1|-|x_{j}-1|).
\label{equ4}
\end{eqnarray}

In Eq.\eqref{equ4}, \textit{j} is the cell index, $ x_{j} $ and $ y_{j} $ are the state variables and the cell outputs, respectively. Each $ a_{j} $ represents a constant parameter and $ i_{j} $ is the threshold value. Also in Eq.\eqref{equ4} $ G_{0} $ and $ G_{s} $ are linear combinations of the outputs and state variables, respectively, of the connected cells. A dynamic model of two generalized CNN cells corresponding to Eq.\eqref{equ4} can be defined as follows:

\begin{eqnarray}
\dot{x}_{1} & = &-x_{1}+a_{1}y_{1}+a_{12}y_{2}+\sum_{k=1}^{2}s_{1k}x_{k}+i_{1}, \nonumber \\
\dot{x}_{2} & = &-x_{2}+a_{21}y_{ 1}+a_{2}y_{2}+\sum_{k=1}^{2}s_{2k}x_{k}+i_{2},
\label{equ5}
\end{eqnarray}

where $ x_{1} $ and $ x_{2} $ are state variables, and $ y_{1} $ and $ y_{2} $ are the corresponding outputs. The MLC circuit equation defined by Eq.\eqref{equ2} can be derived from Eq.\eqref{equ4}, by assuming :	$ x=x_{1} $, $ y=x_{2} $, $ a_{1}=b-a $, $ a_{12} = a_{21}=a_{2}=0$, $ s_{11}=1-b $, $ s_{12}=1 $, $ s_{21}=-\beta $, $ s_{22}=1-\beta(1+\nu) $, $ i_{1}=0 $ and $i_{2} = F(t) $.

Consequently, from Eq.\eqref{equ4} the SC-CNN based MLC circuit model is organized as below : 
\begin{eqnarray}
\dot{x}_{1}&=&-x_{1}+a_{1}y_{1}+s_{11}x_{1}+s_{12}x_{2}, \nonumber \\
\dot{x}_{2}&=&-x_{2}+s_{21}x_{1}+s_{22}x_{2}+F(t),   \nonumber \\
y_{1}&=&0.5 \times(|x_{1}+1|)-(|x_{1}-1|). 
\label{equ8}
\end{eqnarray}
Here $ F(t)= E+\sqrt{D}\xi(t)+ I + f sin (\omega t) $  and $ I=I_{1}+I_{2} $, corresponding to two logic inputs, $ E $ is the bias and the remaining parameters are fixed at $a = -1.02$, $b = -0.55$, $\nu = 0.015$, $\beta = 1.0$ and  $\omega = 1.0$. The MLC circuit in the framework of SC-CNN has been studied numerically, experimentally and analytically in ref.\cite {gunay2010mlc}. \textcolor{black}{ The standard MLC circuit (Fig.\ref{fig19}) consists of a nonlinear resistor which has the three-segment piece-wise characteristics of Chua's diode, a linear resistor, a linear inductor, and a linear capacitor with sinusoidal voltage source. On the other hand the SC-CNN model of MLC circuit (see Fig.\ref{fig3}) mainly consists of a few OP-AMPs along with RC based circuits. Thus the SC-CNN model of the MLC circuit is an inductor-less purely RC based circuit, which leads to the realization of hardware easily. In the present paper, we consider this circuit and investigate the effect of two aperiodic and logical square wave signals on the SC-CNN of MLC circuit. }

\section{Experimental realization of CNN cell of MLC}
\label{sec4}
Fig.\ref{fig3} represents the complete experimental setup for two SC-CNN cells which generate the dynamics of the SC-CNN based MLC circuit. It is clearly obvious from Fig.\ref{fig3}, that the circuit consists of a CNN with two cells, corresponding to the dynamical variables $x_{1} \& x_{2}$ in Eq.\eqref{equ8}. The two state variables $x_{1}$ and $x_{2}$ of Eq\eqref{equ8} are associated with the voltages across the two capacitors $C_{1}$ and $C_{2}$, respectively. The nonlinearity is simply implemented by taking into account the saturation of op-amps.
The circuit realization of the SC-CNN based MLC circuit is achieved with the cell components: $ R_{11}=207K\Omega, \:R_{12}=66K \Omega, \:R_{13}=100K \Omega, \:R_{14}=100K \Omega, \:R_{15}=1K \Omega, \:R_{16}=100K \Omega, \:R_{17}=100K \Omega, \:R_{N1}=220K \Omega, \:R_{N2}=3M \Omega, \:R_{N3}=180K \Omega, \:R_{N4}=16K \Omega, \:R_{21}=100K \Omega, \:R_{22}=6666.6K \Omega, \:R_{23}=100K \Omega, \:R_{24}=100K \Omega, \:R_{25}=1K \Omega, \:R_{26}=100K \Omega, \:R_{27}=100K \Omega, R_{31}-R_{38}=10 K\Omega,~ C_{1}=10nF,\: C_{2}=10nF, \:$ and active element $IC741$ type voltage op-amps with $\pm 12 V$ supply voltages. All the experimental results are obtained by using $ \textit{Agilent (33220A)} $ function generators and $ \textit{Agilent digital oscilloscope (DSO 7014B)} $. The amplitude of the forcing signal $f$ is used to study the dynamics of the CNN based model.\\

\section{Experimental and Numerical results}
\label{sec5}
\subsection{Three level logic input}

Now, we analyze the response of the system \eqref{equ8} to deterministic logic input signal $ I $. Specifically, we drive the system \eqref{equ8} with a low/moderate amplitude signal $I=I_{1}+I_{2}$ with two square waves of strengths $I_1$ and $I_2$ encoding two logic inputs. The inputs can be either 0 or 1, giving rise to four distinct logic input  sets $(I_1,I_2):(0,0),(0,1),(1,0)$ and $(1,1)$. For a logical  `$0$', we set $I_{1}=I_{2}=-\Delta$, whereas for a `$1$', we set  $I_{1}=I_{2}= +\Delta$, where $\Delta$ represents a small/moderate  intensity input signal. As $I=I_{1}+I_{2}$, the input signal sets (0,1) and (1,0) provide the same input signal $ I $. As a consequence, the input signals take the values $-2\Delta,~ 0,$ and $ +2\Delta$ corresponding to the input sets $(0,0),~(0,1)$ or $(1,0),$ and $(1,1)$, respectively.

\begin{figure}[h!]
	\centering	
	\includegraphics[width=0.8\linewidth]{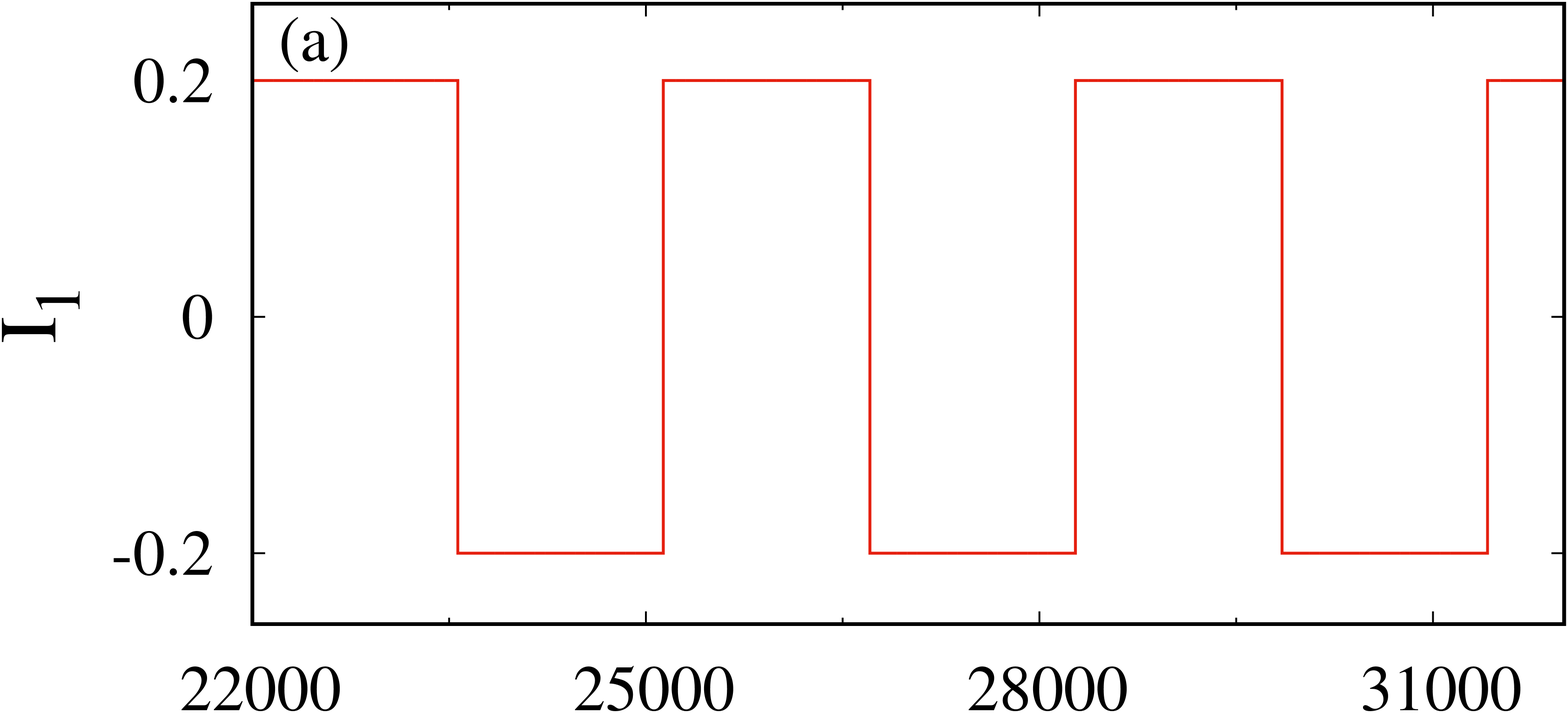} 		
	\includegraphics[width=0.8\linewidth]{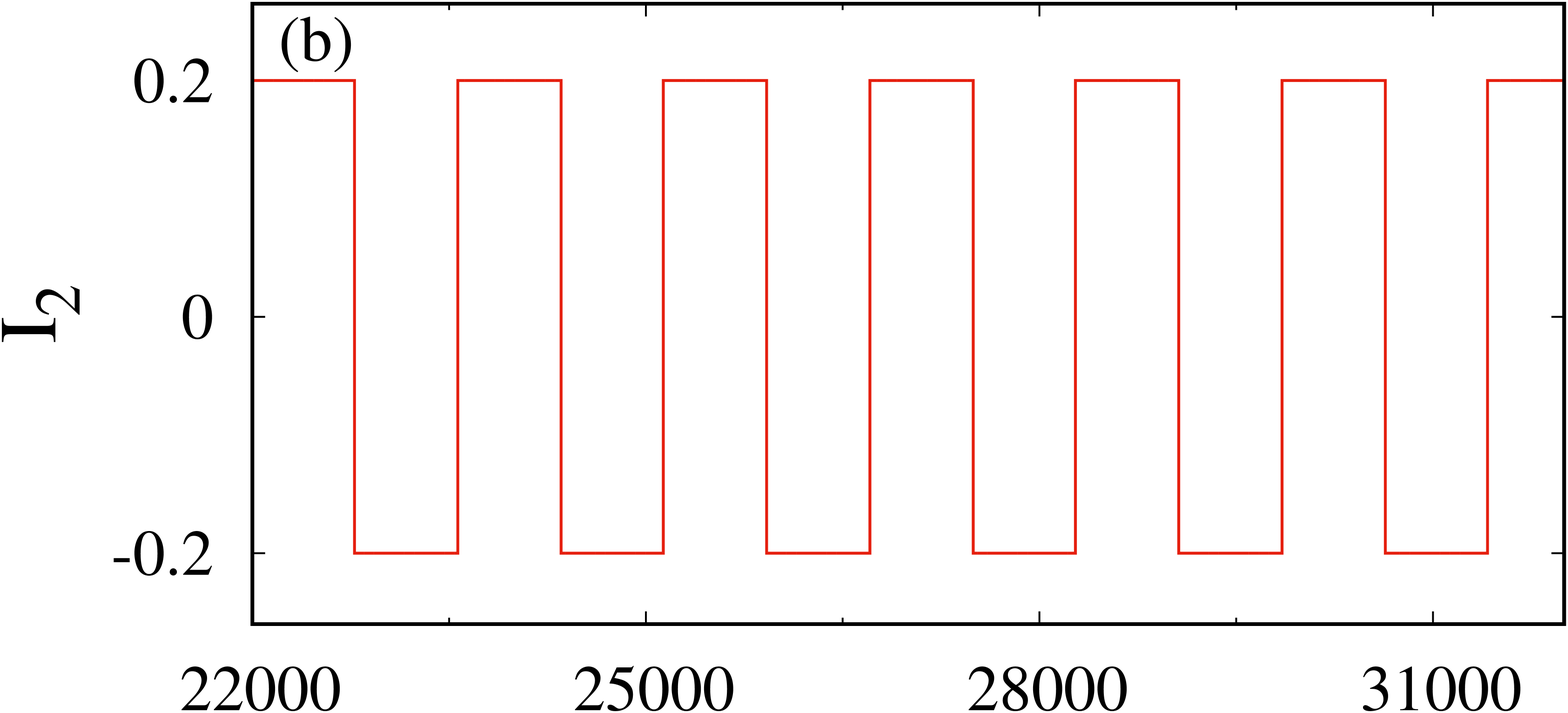} 
	\includegraphics[width=0.8\linewidth]{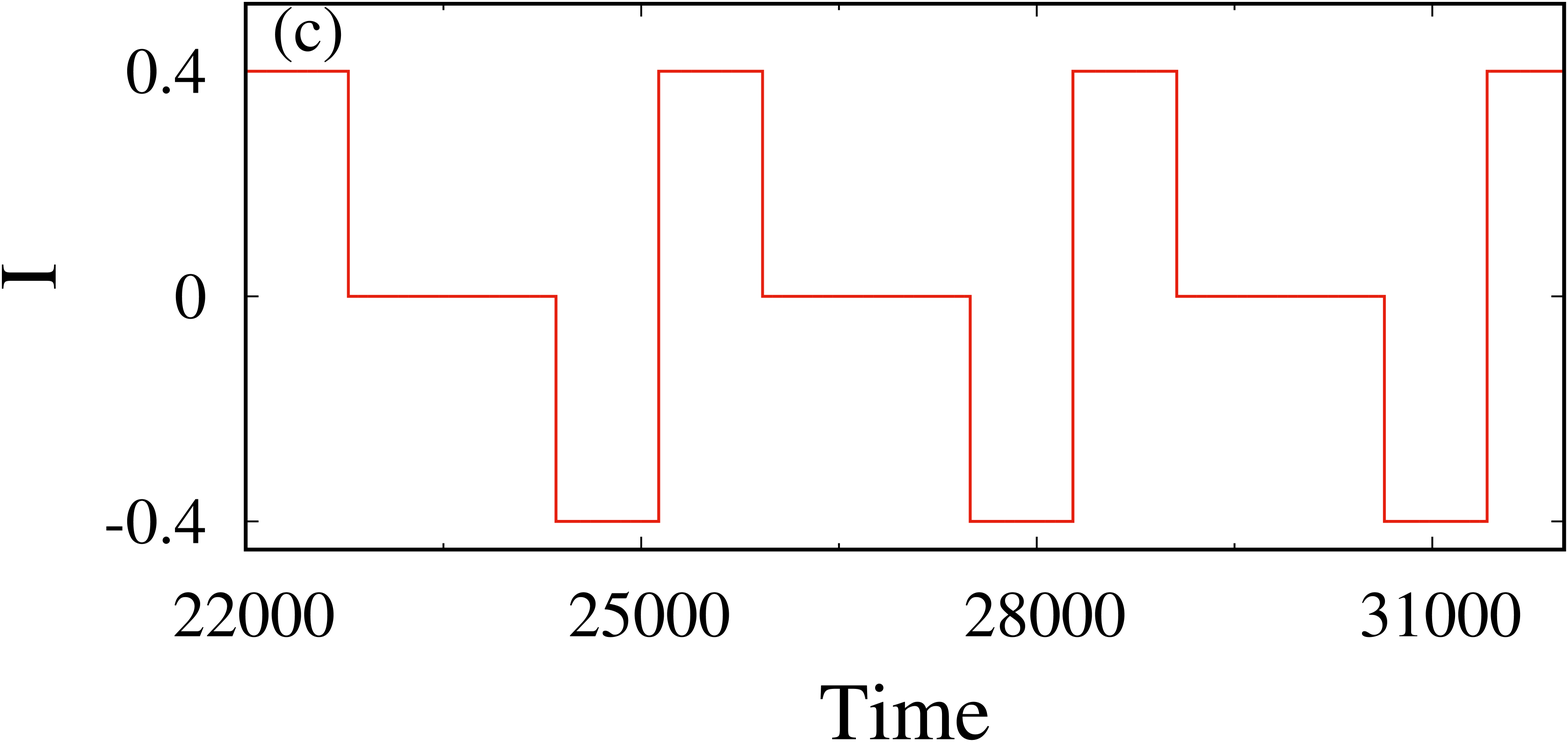}
	\caption{Panels: (a)-(b) correspond to the two different logic inputs of $I_{1}$ and $I_{2}$, respectively. Panel (c) shows a combination of two input signals $I_{1}+I_{2}$. Input $I_{1}=I_{2}=-0.2$ when the logic input is $'0'$ and $I_{1}=I_{2}=+0.2$ when the logic input is $'1'$. The '3' level square waves with $ -0.4 $ corresponding to the input set $ (0,0) $, $ 0 $ for $ (0,1)/(1,0) $ set and $ +0.4 $ for $ (1,1) $ input set.}
	\label{fig7}
\end{figure}
For example, if $\Delta=0.2$ then both the inputs $I_{1}=I_{2}=-0.2$ for logical input `$0$' and values $0.2$ when it is `$1$'. Figs.\ref{fig7}(a) \& \ref{fig7}(b) correspond to input signals $ I_{1}$ \& I$_{2} $, Fig.\ref{fig7}(c) is a three-level square wave. Here, as $ I=I_{1}+I_{2} $, the amplitude value $ -0.4 $ of the wave is for input sets $ (0,0) $, 0 for input sets $ (0,1)  $ or $ (1,0) $ and $ +0.4 $ for the input set $ (1,1) $. Then the output of the appropriate logic gate is determined by the well in which the dynamical variable $ x_1 $ or $ x_2 $ resides. More specifically, for the OR/NOR and AND/NAND logic operations, the state variables hop between the left and right/center wells depending on the input streams, whereas for the XOR/XNOR logic gates, the state variables leap among all the three wells equally.

\begin{figure}[!h]
	\centering
	\includegraphics[width=0.8\linewidth]{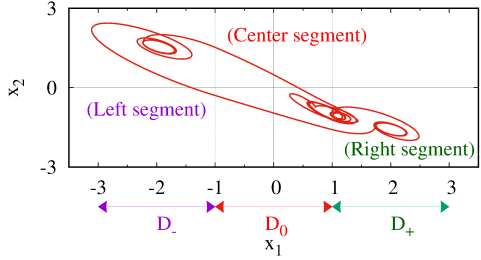}
	\caption{Phase-space of attractor of Eq.\ref{equ8} for OR logic behavior. It is having three segments: Left segment $ D_{-} $ ($ x_{1}<-1 $), Center segment $ D_{0} $($ -1 \leq x_{1} \leq 1 $) and Right segment $ D_{+} $($ x_{1}>+1 $)}
	\label{fig17}
\end{figure}

\subsection{Realization of OR/NOR and AND/NAND logic gates}

Two typical logical operations in digital circuits are OR and AND or the complementary gates NOR and NAND. To realize these gates, at least two inputs need to be converted into a single output. For example, in the case of OR logic operation, at least one of the two inputs is an 'ON' state, so as to get an 'ON' output, while for AND logic operation, it is essential to be in the 'ON' for both the inputs, so as to obtain an 'ON' output(see Table.\ref{Tab2}). The logical response of a desired type can be extracted by finding the solution of state variables $x_{1}$ and $x_{2}$ from \eqref{equ8}. One can define or extract an appropriate gate by considering the state variables $x_{1}$ and $x_{2}$ in the phase space. For example, in our system \eqref{equ8}, the system oscillates in three regions: namely 1) $D_{-}$ region - the state variable $x_{1}$ resides in the region $x_{1}<-1$, 2) $D_{0}$ region - the state variable $x_{1}$ oscillates between $-1\leq x_{1} \leq +1$ and 3) $D_{+}$ region - the state variable $x_{1}$ exists for $x_{1}>1$.

\begin{figure}[!h]
	\centering
	\includegraphics[width=0.7\linewidth]{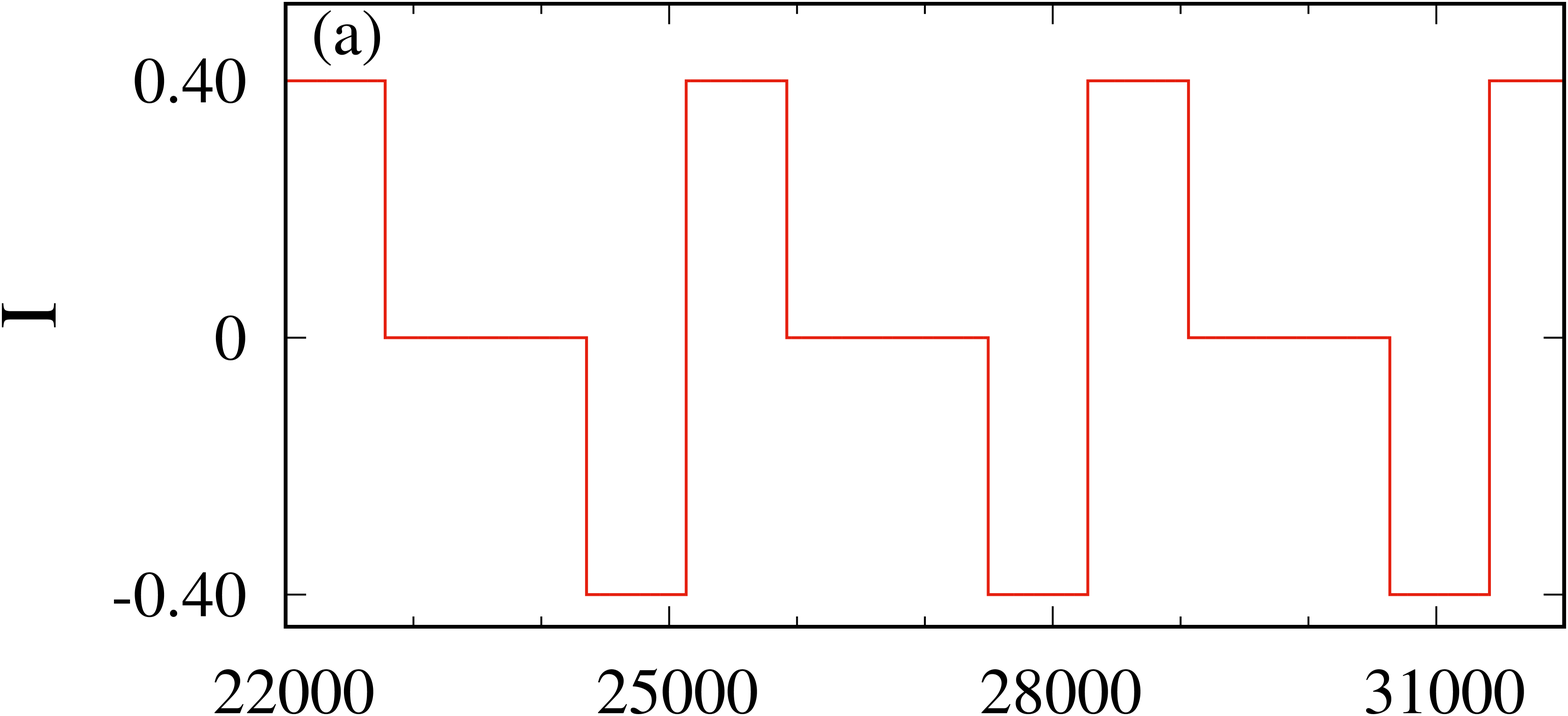}
	\includegraphics[width=0.7\linewidth]{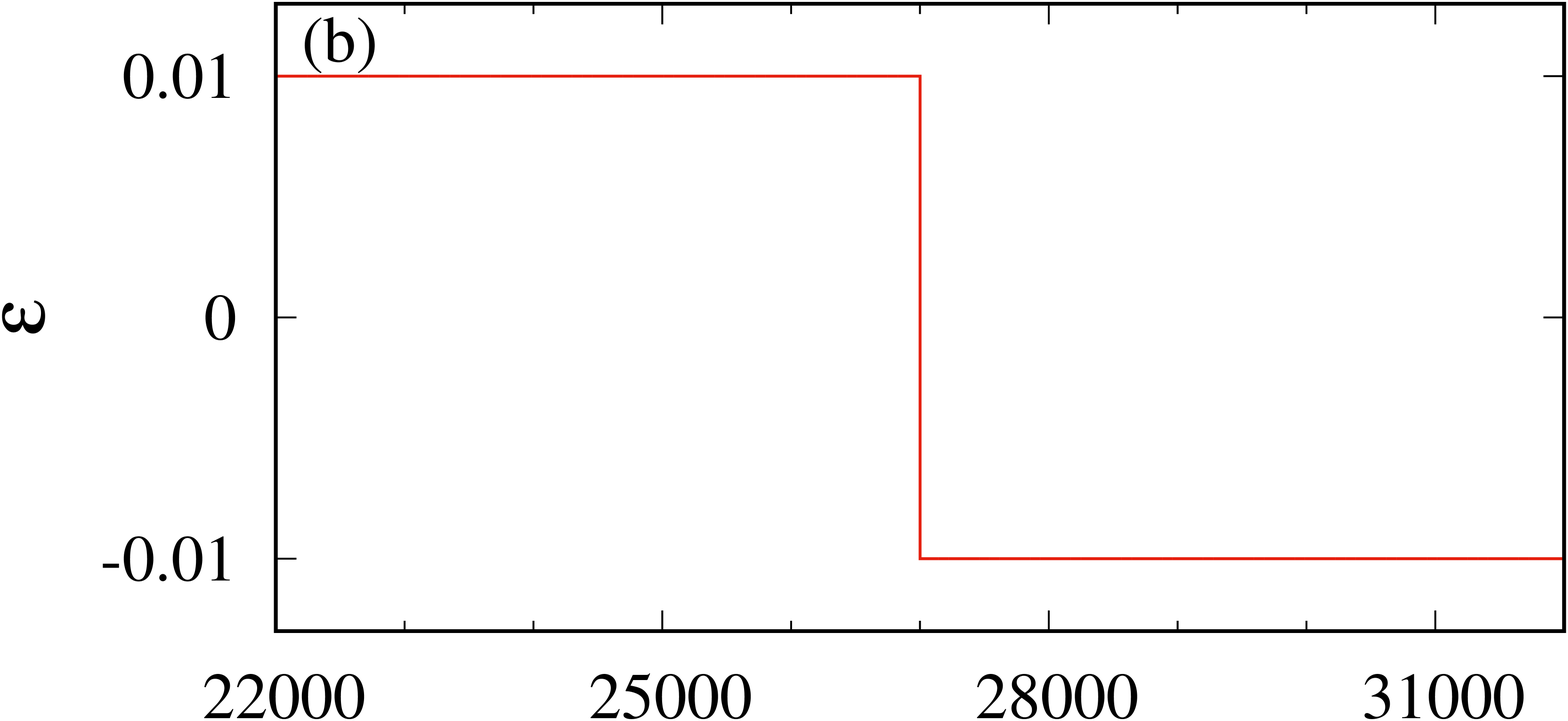}
	\includegraphics[width=0.7\linewidth]{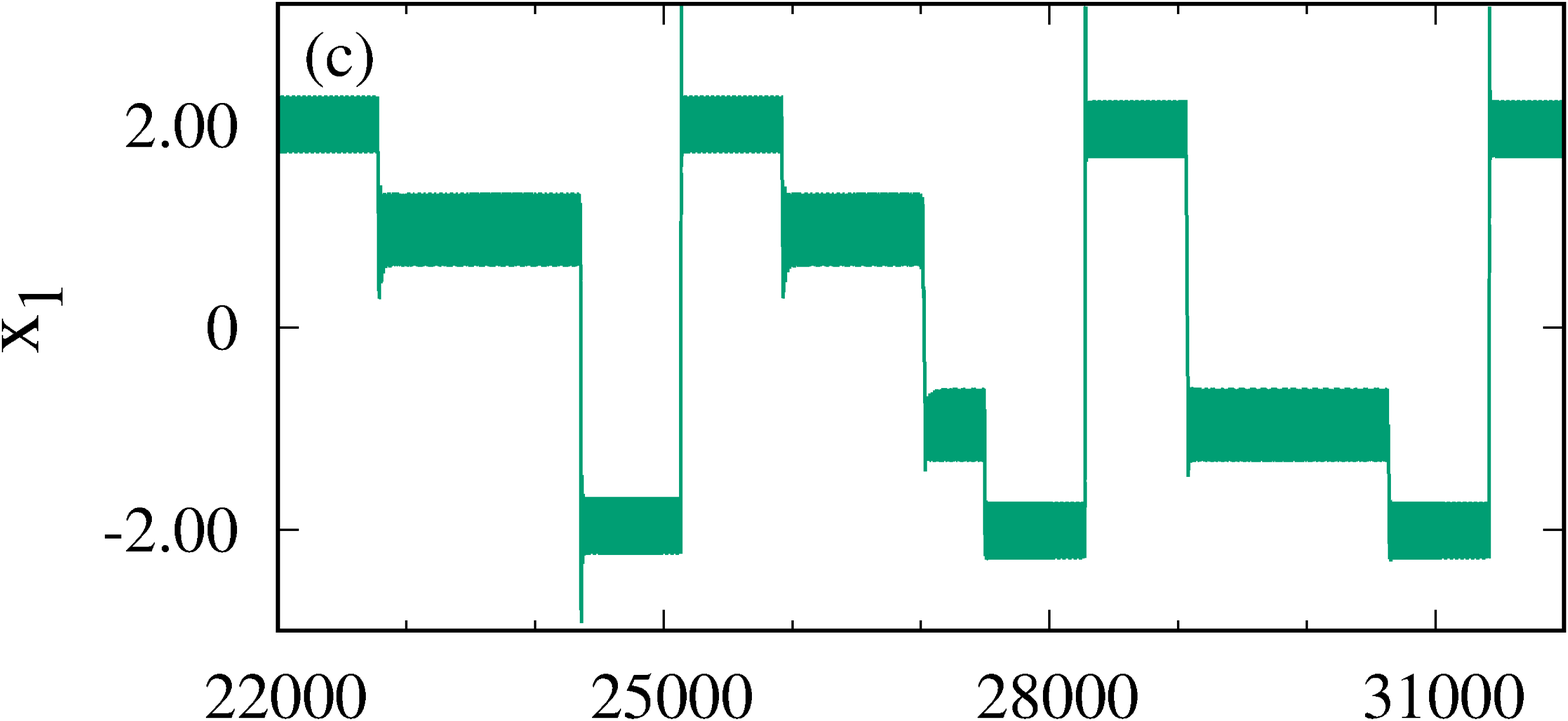}
	\includegraphics[width=0.7\linewidth]{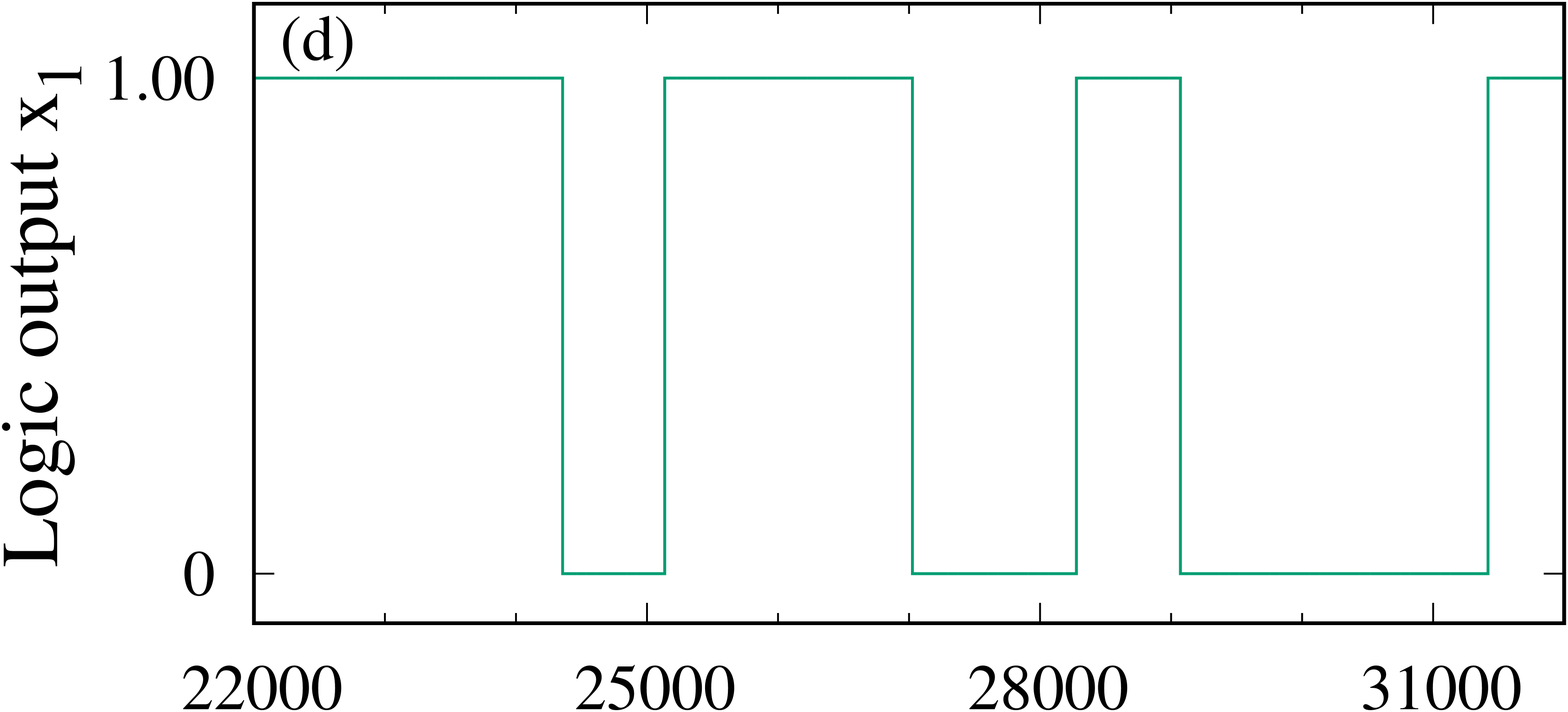}
	\includegraphics[width=0.7\linewidth]{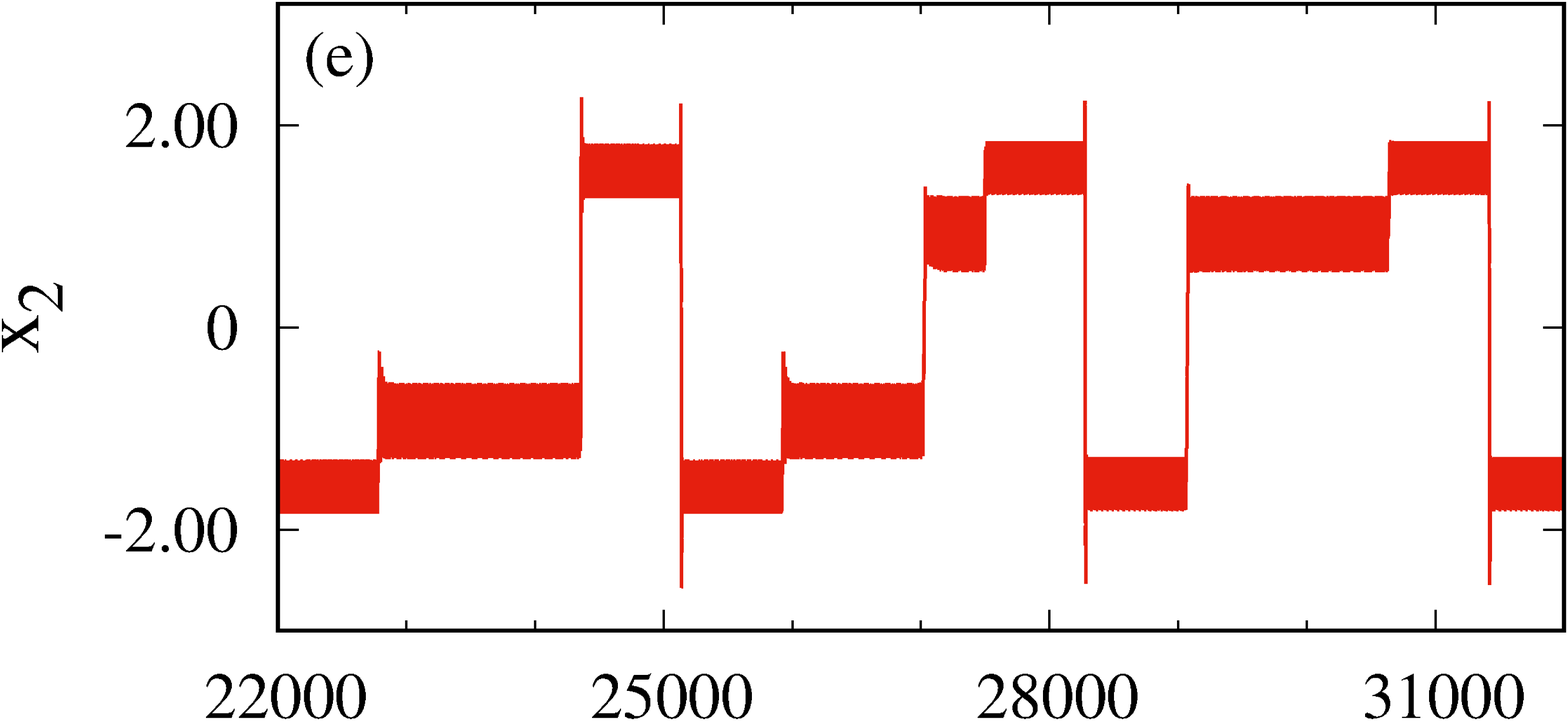}
	\includegraphics[width=0.7\linewidth]{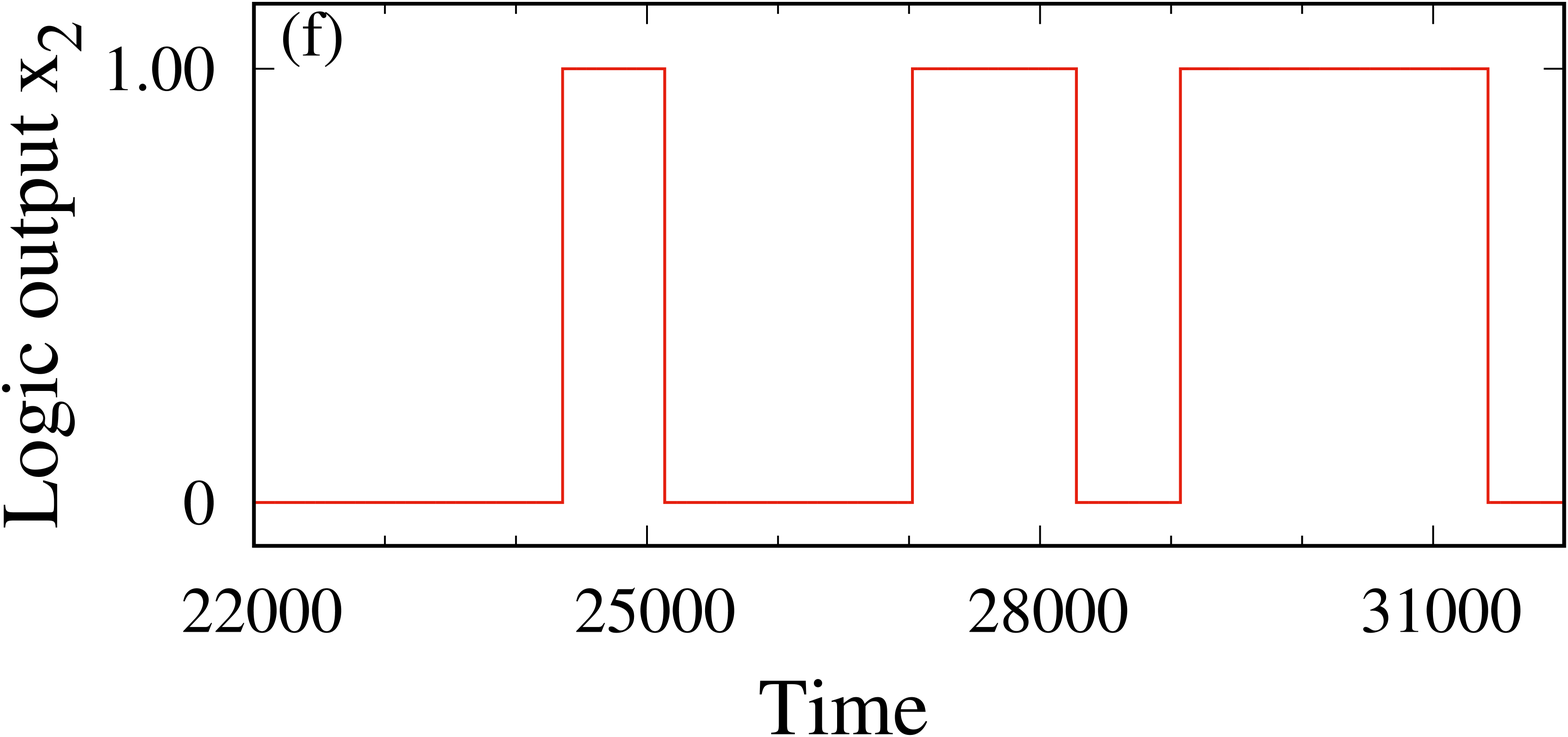}
	\caption{Panel (a) shows a combination of two input signals $I=I_{1}+I_{2}$. Input $I_{1}+I_{2}=-0.4$ when the logic input is $'0'$ and $I_{1}+I_{2}=+0.4$ when the logic input is $'1'$. Panel (b) is the constant bias, when it varies from $ 0.01 $ to $ -0.01 $. Panels (c) and (e) represent the corresponding dynamical responses OR/AND and NOR/NAND logic gates of the system $ x_{1}(t) $ and $ x_{2}(t) $, respectively, under periodic forcing with $f=0.1$. \textcolor{black}{Panels (d) and (f) are the corresponding logic outputs of panels (c)  and (e), respectively.}}
	\label{fig10}
\end{figure}

Next to obtain the OR logic operation, when the state variable $x_{1}$ is in the $D_{-}$ region of the system \eqref{equ8}, the output state is assumed to be the logical output value '0' and when the state variable $x_{1}$ resides in any one of the regions $D_{0}$ and $D_{+}$, the output is assumed to be the logical value '1'. With this input-output association, we solve equation \eqref{equ8} numerically by fixing the parameters as $I_{1} = I_{2} =0.2$, $E = 0.01, f = 0.1$ and the remaining parameter values are assigned the same values as given in section \ref{sec3}. 

Now, we analyze the attractors of the system \eqref{equ8}. In Fig.\ref{fig17}, it is obvious that for input signal $ I $ which corresponds to either (1,1) or (0,1)/(1,0) states, the response of the system `$ x_1 $' hops either in the $ D_+ $ region or in the positive side of $ D_0 $ region of the phase space and when the input signal is in the (0,0) state, the value of the state variable `$ x_1 $', is in other segment, namely the $ D_- $ region. The logical output is `$ 1 $' when $ x_{1}(t) > 0 $ and it is `0' for $ x_{1}(t)<0 $. As a consequence, for the two sets of input streams (1,1) and (0,1)/(1,0), the attractor of the system is bounded in $ x_{1}(t) > 0 $ region and its is $ x_{1}(t) < 0 $ region for the (0,0) state. This confirms the fact that the dynamical attractor is although chaotic, it behaves as the logical OR operation. It is also interesting to note in Fig.\ref{fig17} that when $ x_{1}(t) > 0 $, the $ x_2(t)$ variable of the attractor is bounded in the region $ x_2(t)<0 $. That is the response of the $ x_{2} $ variable turns out to be the inverted output of the $ x_{1} $ variable. As a result, these two variables $ x_{1}(t) $ and $ x_{2}(t) $ produce logic gates OR and NOR in parallel in this circuit. 

These results are clearly illustrated in the time trajectory plot as shown in Figs.\ref{fig10}. In Fig.\ref{fig10}(a) the three level diagram for the logic input $ I=I_{1}+I_{2} $ is shown and as the bias value `$E$' is changed from $ 0.01 $ to $ -0.01 $ (see Fig.\ref{fig10}(b)) the response of the system morphs from OR gate to the other logical AND gate in the $ x_{1}(t) $ regime of Fig.\ref{fig10}(c) and the corresponding logic output as shown in Fig.\ref{fig10}(d). For this case, it is found that for the (1,1) state, the attractor is in the $ D_{+} $ region of phase space while for (0,1) /(1,0) and (0,0) state it is either in the negative $ D_{0} $ region or in the $ D_{-} $ region. Hence, switching the bias value from +0.01 to -0.01 morphs logical OR gate to logical AND gate (see Figs.\ref{fig10}(c) \& \ref{fig10}(d)). In a similar fashion the other state variable $ x_{2}(t) $ mimics the inverted output signal of $ x_{1}(t)$, thereby produces a clean logical NOR for bias value $ E = 0.01 $. It is further observed that the response of state variable $ x_{2}(t) $ of the system morphs from NOR logic to NAND logic (see Figs.\ref{fig10}(e) \& \ref{fig10}(f)) when the threshold switches from the value of 0.01 to -0.01. Thus the system \eqref{equ8} produces logic OR/AND through one of state variable $ x_{1} $ while its complementary logic function NOR/NAND is realized via the other state variable $ x_{2} $. 

\begin{figure}[!h]
	\centering
	\includegraphics[width=1.0\linewidth]{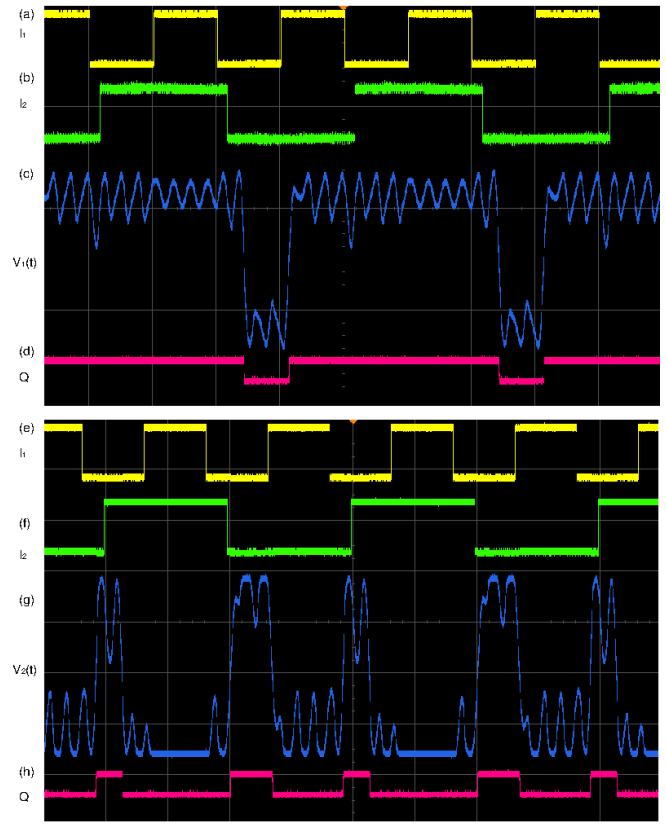} 
	\caption{Realization of the OR/NOR logic gates in experimental electronic circuits. Panels (a) \& (e) and (b) \& (f) are the inputs $ I_{1} $ and $ I_{2} $ (if the logic input is 0, it corresponds to $ -100mV $, and $ +100mV $ is considered as logic input 1). \textcolor{black}{Panels (c) \& (d)  clearly indicate the OR logic output (when $ v_{1}(t)>0 $ it is considered as logic input 1, and $ v_{1}(t)<0 $ its logic input is considered as 0).  Panels (g) \& (h) yield the complementary NOR logic gate. The bias voltage is fixed as $ +10 mV $ and the experimental output was observed in Agilent DSO 7014B. }}
	\label{fig11}
\end{figure}

\begin{figure}[!h]
	\centering
	\includegraphics[width=1.0\linewidth]{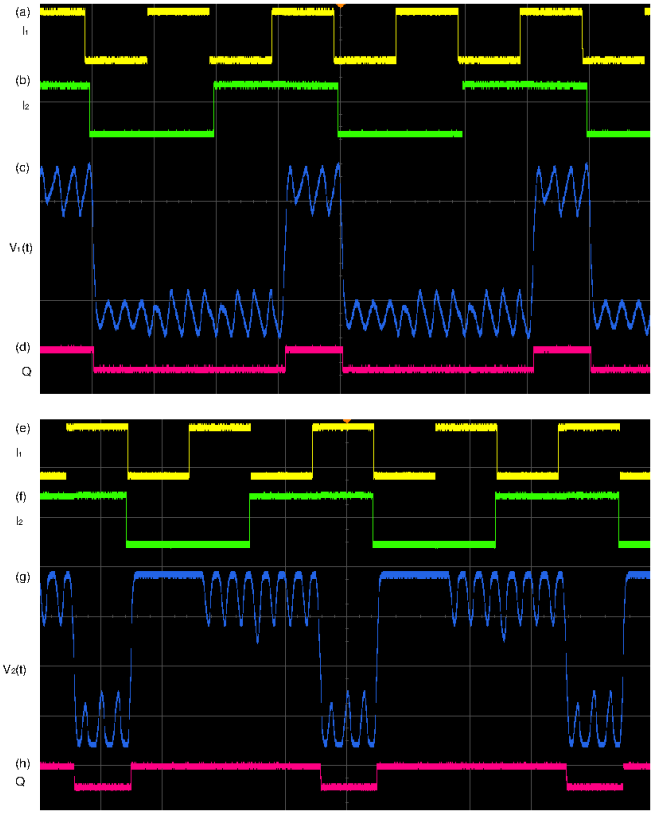}
	\caption{Realization of the AND/NAND logic gates in experimental electronic circuits. \textcolor{black}{ Panels (a) \& (e) and (b) \& (f) are the inputs $ I_{1} $ and $ I_{2} $ (if the logic input is 0, it corresponds to $ -100mV $, and $ +100mV $ is considered as logic input 1). Panels (c) \& (d)  clearly indicate  the AND logic output (when $ v_{1}(t)>0 $ it is considered as logic input 1, and $ v_{1}(t)<0 $ is considered as the input 0). Panels (g) \& (h) yield the complementary NAND logic gate. The bias voltage is fixed as $ -10 mV $ and the experimental output was observed in Agilent DSO 7014B. }}
	\label{fig12}
\end{figure}

\begin{figure}[!h]
	\centering
	\includegraphics[width=0.9\linewidth]{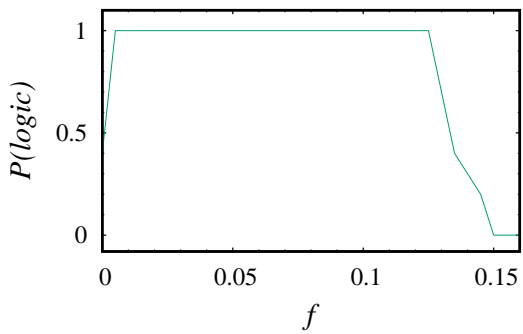}
	\caption{The probability distribution of obtaining logical behavior for different values of $f$ with  bias value $ E=0.01 $ for OR/NOR gates and $ E=-0.01 $ for AND/NAND gates.}
	\label{fig23}
\end{figure}

Next we verify the numerical results discussed earlier in the electronic circuit analog of the nonlinear system described by Eq.\eqref{equ8}  and ascertain its robustness in experiments. The complete circuit realization for the two SC-CNN cells generating the dynamics of SC-CNN MLC circuit is shown in Fig.\ref{fig3}. As pointed out earlier, Fig.\ref{fig3} consists of two cells, each corresponding to one dynamical variable in Eqs.\eqref{equ8}. The two state variables $ x_{1} $ and $ x_{2} $ of Eqs.\eqref{equ8} are associated with the voltages $ v_{1} $ and $ v_{2} $ across the two capacitors $ C_{1} $ and $ C_{2} $, respectively. Now we fix $ I_{1}= 100mV $, $ I_{2}= 100mV $ and $ bias= +10mV $ corresponding to dimensionless units of the circuit parameter that we discussed earlier. The changes in the dynamics of the circuit under the effect of input streams are obtained by measuring the voltages $ v_{1} $ and $ v_{2} $ across the capacitors $ C_{1} $ and $ C_{2} $, respectively. The logic input signals $ I_{1} $ and $ I_{2} $ are $ +100mV $ when the logic input is `1' and $ -100mV $ for `0' bias voltage $ bias=10mV $. For this case, the response of the circuit exhibits OR gate when we measure the voltage $ v_{1} $ across $ C_{1} $ and NOR gate when we measure the voltage $ v_{2} $ across $ C_{2} $ (see Fig.\ref{fig11}). In a similar way, for $bias=-10mV$, the system gives the AND gate when the voltage $ v_1 $ is measured across $ C_{1} $ and NAND gate when the voltage $ v_{2} $ is measured across $ C_{2} $ (see Fig.\ref{fig12}). 
\textcolor{black}{The logical output `$ Q $' in Figs.\ref{fig11}(d)/\ref{fig11}(h) \& \ref{fig12}(d)/\ref{fig12}(h) are obtained as $ v_{1}/v_{2} $ by feeding the output response to an appropriate comparator circuit \cite{campos2010simple}. Further, we can quantify the process of obtaining a given logic output by calculating $ P(logic) $ as shown in Fig.\ref{fig23} in which it is observed that the fundamental logic operation OR is realized in an optimal width of forcing amplitude `$ f $'. It is interesting to note that the logic operation is realized for subthreshold input which results in an optimal window of forcing value where $ P(logic)$ tends to $ 1$.}

\subsection{Realization of XOR and XNOR logic gates}

\begin{figure}[!h]
	\centering
	\includegraphics[width=0.8\linewidth]{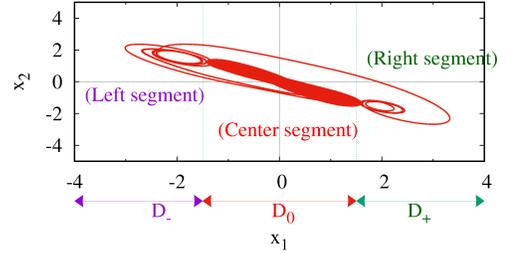}
	\caption{Figure shows the phase-space section of XOR logic behavior. It is having three segments: Left segment $ D_{-} $ ($ x_{1}<-1.5 $), Center segment $ D_{0} $($ -1.5 \leq x_{1} \leq 1.5 $) and Right segment $ D_{+} $($ x_{1}>+1.5 $).}
	\label{fig22}
\end{figure}

The logic gate XOR is quite different from the previously discussed gates AND, OR, NAND and NOR. XOR gates admit a logic output of `1' or high logic level if the inputs are at different logic levels, that is either of 0 and 1 or 1 and 0. Conversely, the output will be a `0' or a low logic if the inputs are at the same logic levels 0 and 0 or 1 and 1. The function of XOR gate is to start with a regular `OR' gate. However, the output is inhibited from going to `1' or high when both the inputs are high or `1' and it takes logic low or `0' even when both the inputs are high or `1'. For these conditions, a bistable nonlinear system fails to satisfy them. Since the system \eqref{equ8} has a three segment piece-wise continuous function, it results in the gradient of a triple well energy potential. As a result, the present system \eqref{equ8} has the ability to produce XOR and XNOR gates, apart from producing the fundamental logic gates OR,AND, NOR and NAND. To realize XOR, we set the output to be logical `1' if the state variable $x_{1}$ lies in between $-1.5\leq x_{1} \leq +1.5$ and the output is assumed to be logic `0' if the state variable $x_{1}$ resides anywhere else, that is, when the attractor of system \eqref{equ8} resides in the $D_{0}$ region, the response output is assumed to be high or '1' and '0' otherwise, when the attractor may be in the $D_{+}$ region or $D_{-}$ region (see Fig.\ref{fig22}).

\begin{figure}[!h]
	\centering
	\includegraphics[width=0.8\linewidth]{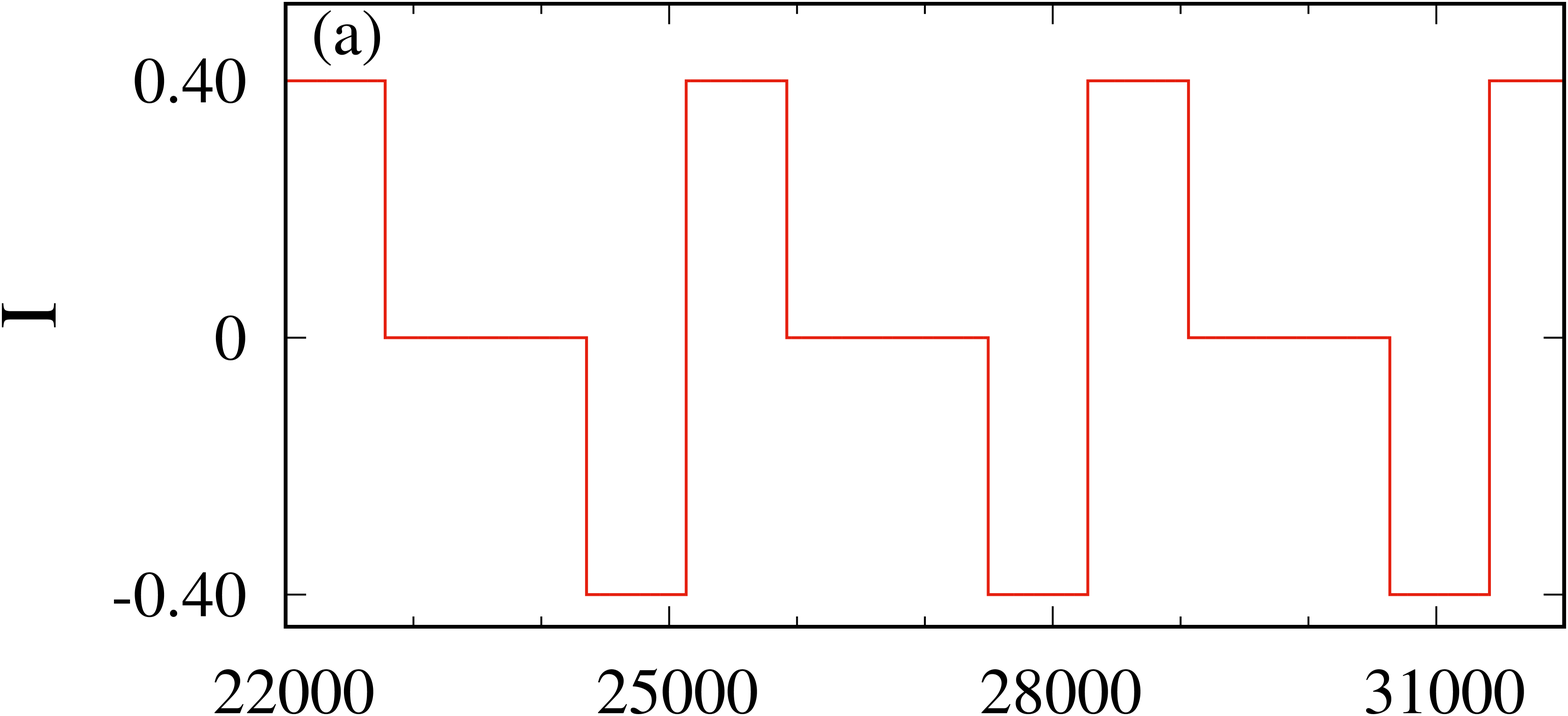}
	\includegraphics[width=0.8\linewidth]{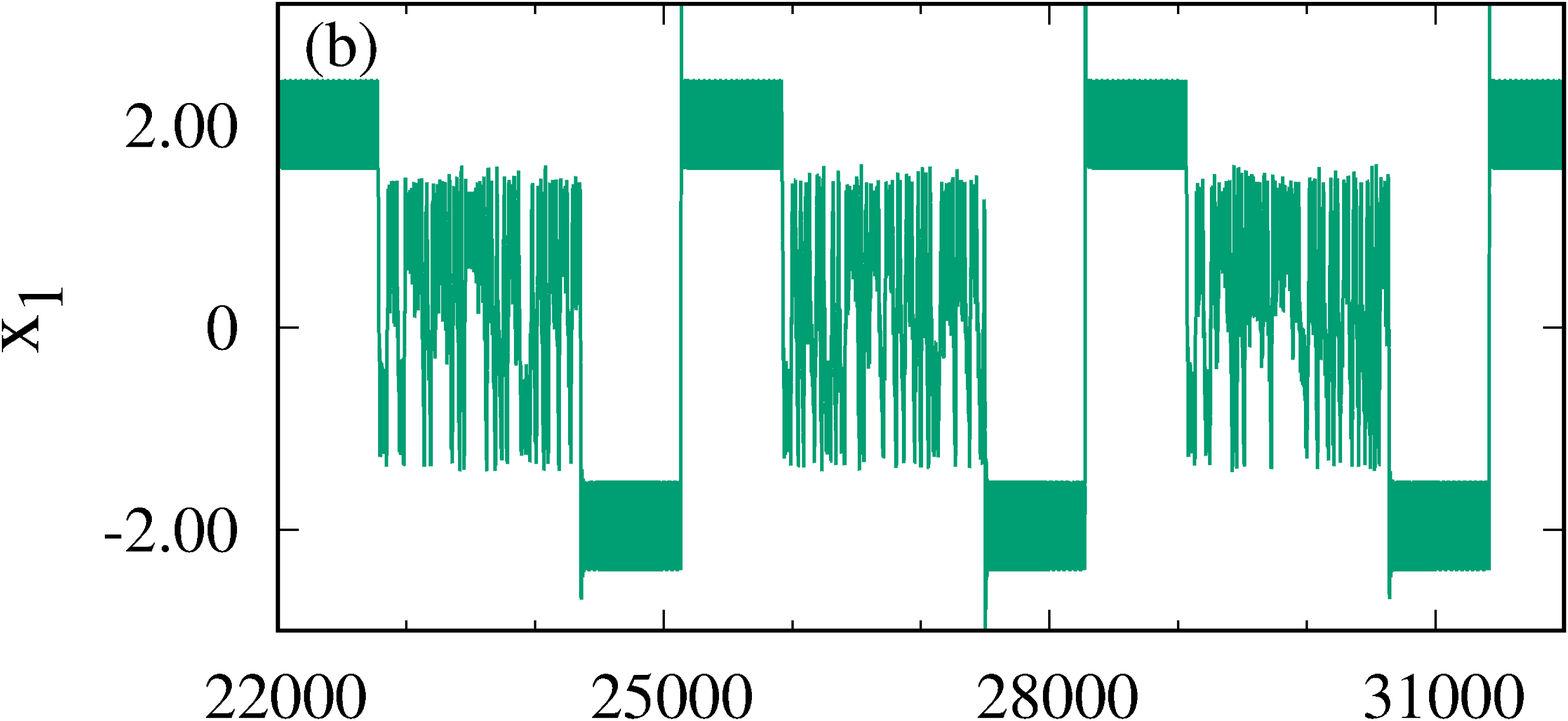}
	\includegraphics[width=0.8\linewidth]{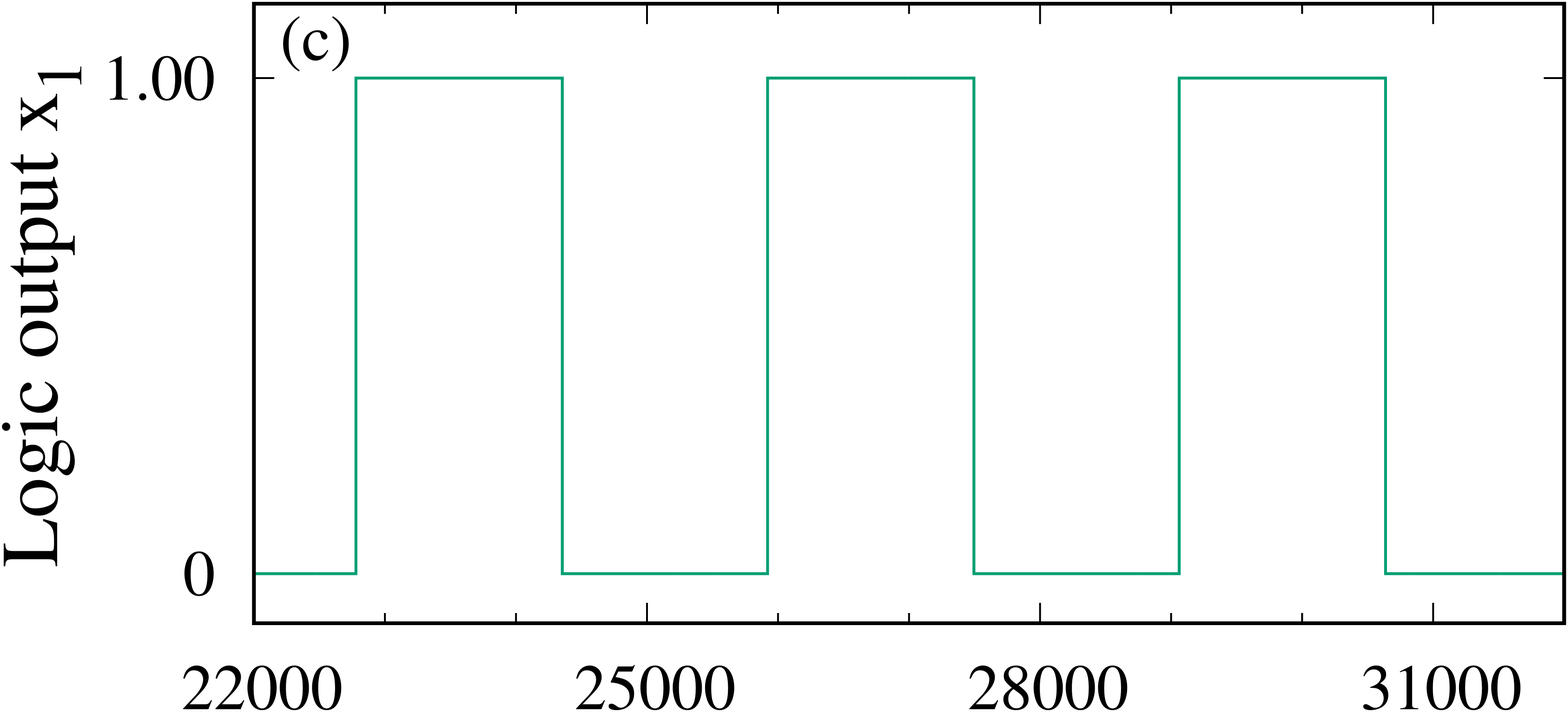}
	\includegraphics[width=0.8\linewidth]{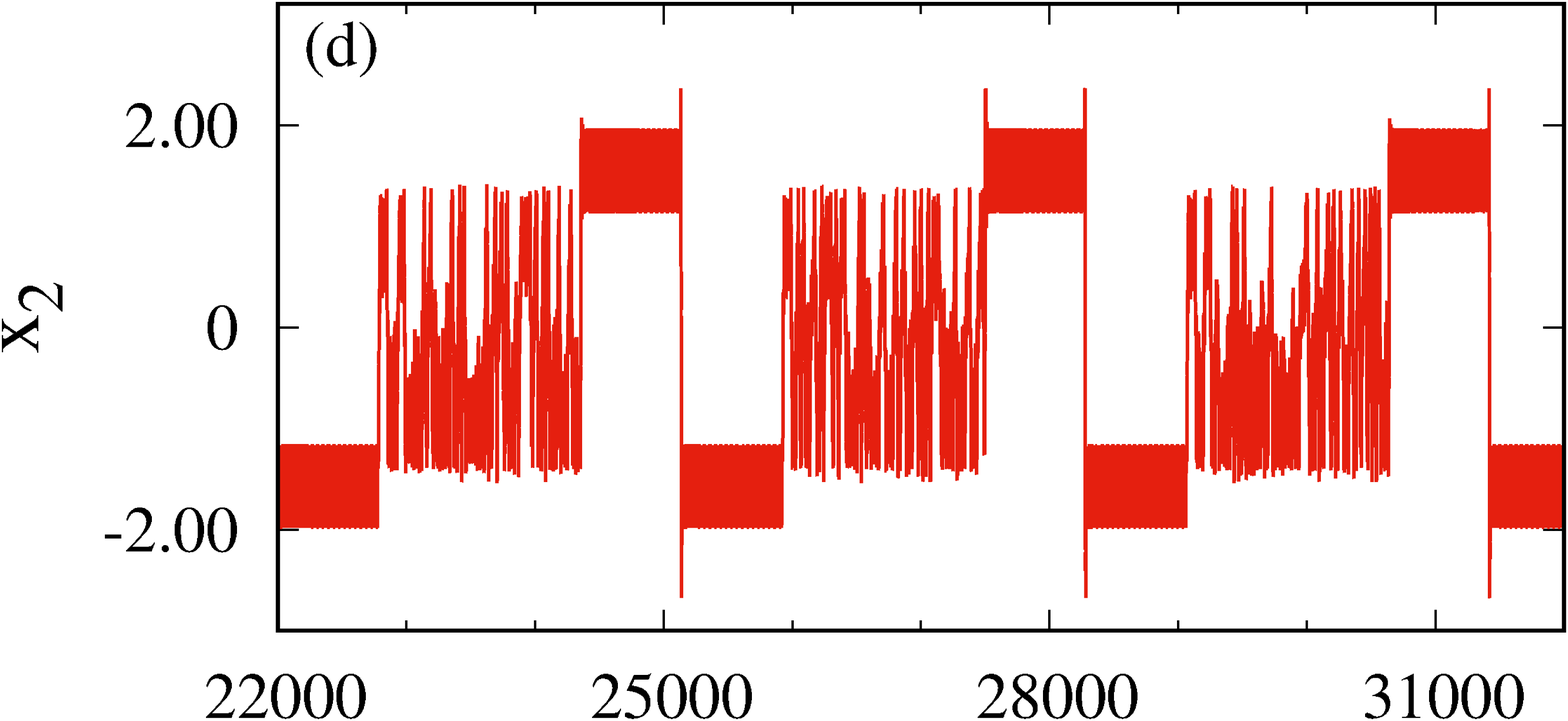}
	\includegraphics[width=0.8\linewidth]{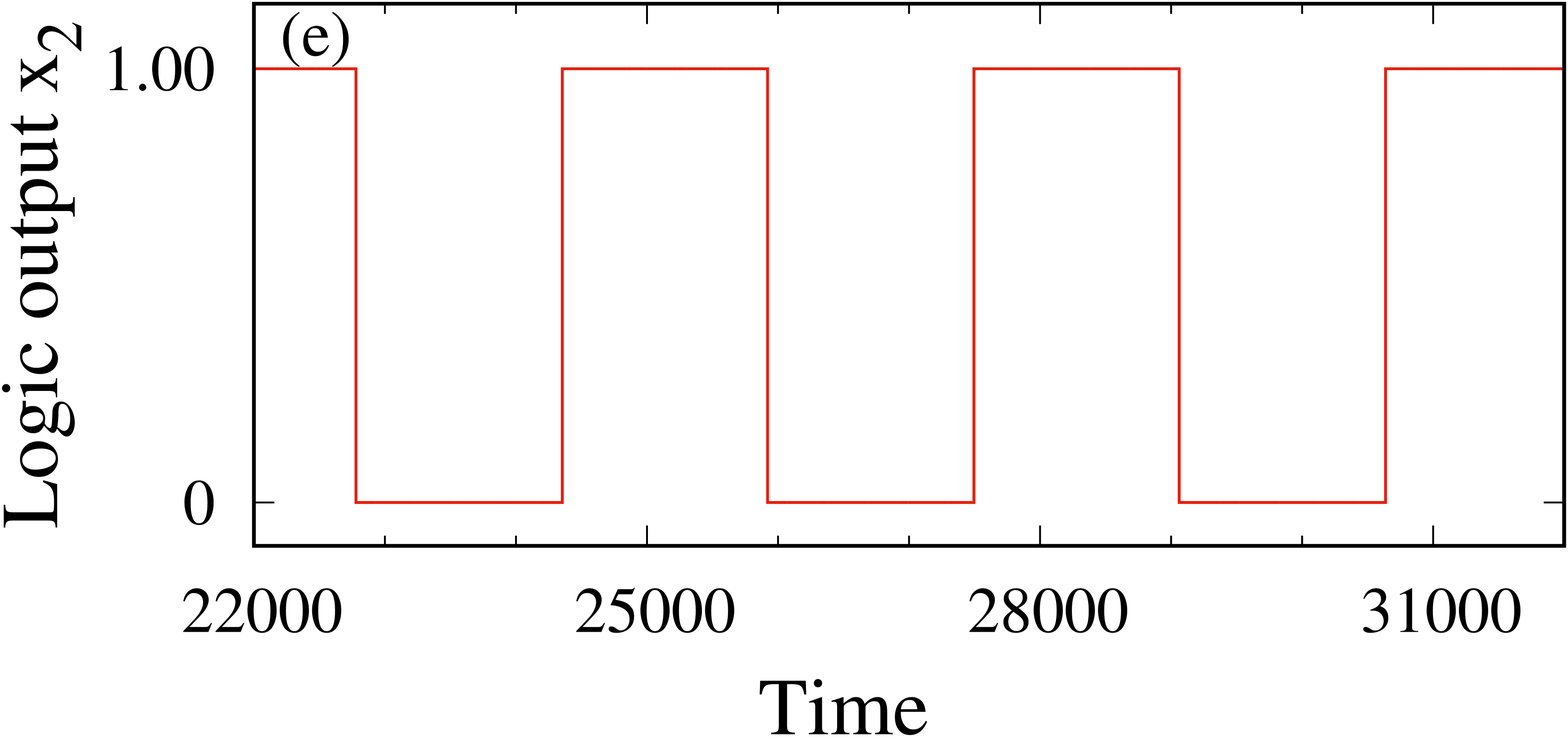}
	\caption{Panel (a) shows a combination of two input signals $I_{1}+I_{2}$. Input $I_{1}=I_{2}=-0.2$ when the logic input is $'0'$ and $I_{1}=I_{2}=+0.2$ when the logic input is $'1'$.\textcolor{black}{ Panels (b) \& (d) represent the corresponding dynamical responses in the form of XOR and XNOR logic gates of the system through the variable $ x_{1}(t) $ and $ x_{2}(t) $  under periodic forcing $ f=0.16 $, with fixed bias parameter $ E=0.01 $, and panels (c) \& (d) represent the logic outputs of panels (b) \& (d). }}
	\label{fig8}
\end{figure}

\begin{figure}[!h]
	\centering
	\includegraphics[width=1.0\linewidth]{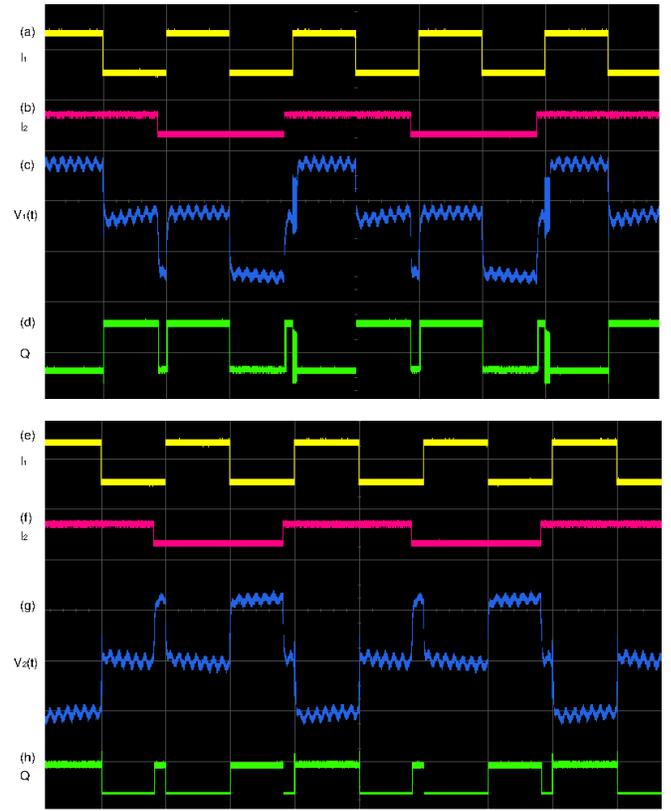}
	\caption{Realization of the XOR/XNOR logic gates in experimental electronic circuits. \textcolor{black}{ Panels (a) \& (e) and (b) \& (f) are the inputs of  $I_{1} $  and $I_{2} $ (if the logic input is 0, it corresponds to $ -100mV $, and $ +100mV $ is considered as logic input 1). Panels (c) \& (d)  clearly indicate the XOR logic outputs (when $ D_{-} \leq v_{1}(t) \geq D_{+} $ is considered as logic input 0, and $ D_{-} \geq v_{1}(t) \leq D_{+} $ its logic input is considered as 1).  Panels (g) \& (h) yield the complementary XNOR logic gate. The bias voltage is fixed as $ +10 mV $ and the experimental output was observed in Agilent DSO 7014B.} }
	\label{fig9}
\end{figure}

With this input-output correspondence, we solve the Eq.\eqref{equ8} numerically by fixing the system parameters as $I_{1}=I_{2}=0.2$, $ E=0.01 $ and $ f=0.16 $. It is clearly shown in the phase space diagram Fig.\ref{fig22} that for the (0,0) state the attractor of the system resides in the $D_{-}$ region of the phase space, for the input (0,1) or (1,0) states, it hops in between -1 to +1 in the $ D_{0} $ region and for (1,1) state it is in the $ D_{+ } $ region. Thus for the state (0,1) or (1,0) the output response of the system is considered to be logical value `1' since the state variable $  x_{1} $ of the system hops in the region $D_{0}$ and for the other states (0,0) or (1,1) the response of the output is considered to be the logical value `0' because the state variable $x_{1}$ oscillates either in the $ D_{-} $ or in $ D_{+} $ region depending on the input streams. Hence, the output response of the system admits a high logic output for inputs of different logic levels, either of 0 and 1 or 1 and 0, and it is a low logic output for the same logic levels, namely 0 and 0 or 1 and 1.\\

Thus for the above set of parameter values, the system \eqref{equ8} admits logic XOR gate. It is also clearly demonstrated in phase diagram (see Fig.\ref{fig22}) that the response of the state variable $x_{2}$ turns out to be the inverted output of the other state variable $x_{1}$. As a consequence, when the state variable $x_{1}$ produces the logical XOR gate, the other state variable $x_{2}$ produces the complementary gate, namely the logical XNOR gate. Thus the present system has the ability to produce XOR gate in one of the state variables $x_{1}$ and XNOR gate in other state variable $x_{2}$ parallely without altering any system parameter. \textcolor{black}{ These results are also illustrated in the time trajectory plots in Fig.\ref{fig8}. In Figs.\ref{fig9}(c), \ref{fig9}(d)\& \ref{fig9}(g), \ref{fig9}(h), it is demonstrated how the response of the system variables $x_{1}$ and $x_{2}$ behave as logical XOR and XNOR operations, respectively, under different input streams as shown in Figs.\ref{fig9}(a),\ref{fig9}(b),\ref{fig9}(e) and \ref{fig9}(f).}\\

\begin{figure}[!h]
	\centering
	\includegraphics[width=0.9\linewidth]{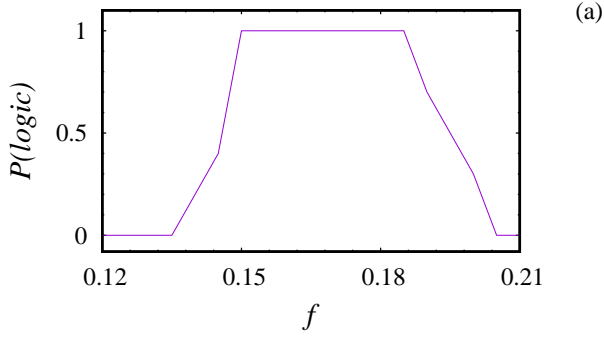}
	\caption{The probability distribution of obtaining logical behavior for different values of $f$ with  bias value $ E=0.01 $ for XOR/XNOR gates.}
	\label{fig24}
\end{figure}

To substantiate our numerical simulations, we carried out the experimental realization of XOR and XNOR gates by using the analog circuit given in Fig.\ref{fig3}. For this purpose, we fix the parameter values as $ I_{1}=-100mV $, $ I_{2}=100mV $ and $bias=+10mV$ corresponding to the dimensionless parameters. The logical XOR and XNOR responses are obtained by measuring the voltages $ v_{1} $ and $ v_{2} $ across the capacitors $ C_{1} $ and $ C_{2} $, respectively. It is obvious from Fig.\ref{fig9} that the signals measured across $ C_{1} $ and $ C_{2} $  behave as logical XOR and logical XNOR operations under different three-level input streams. Thus our experimental study confirms the fact that the considered circuit system has the potential to produce not only the fundamental gates OR/NOR and AND/NAND, but it can also emulate the logical XOR and XNOR operations as well. \textcolor{black}{The logical outputs `$ Q $' in Figs.\ref{fig9}(d)/\ref{fig9}(h) are obtained as $ v_{1}/v_{2} $ by feeding the output response to an appropriate comparator circuit \cite{campos2010simple}. Also, we can quantify the process of obtaining a given logic output by calculating $ P(logic) $ as shown in Fig.\ref{fig24}, where it is observed that the fundamental logic operation XOR is realized in an optimal width of forcing amplitude `$ f $'. It is interesting to note that the logic operations are realized for subthreshold input which results in an optimal window of forcing value where $ P(logic)$ tends to $ 1$. }

\subsection{Set-Reset Memory latch}

\begin{table}
	\begin{center}
		\caption{ Truth Table of Set-Reset (SR) flip-flop}
		\vspace{0.2cm}
		\begin{tabular} {|c| c| c| l |}
			\hline
			 $(I_{1})$ Set & $(I_{2})$ Reset  & Output & State \\
			\hline
			0  & 0 & Q & Last state \\
			\hline
			0  & 1  & 0 & Reset\\
			\hline
			1  & 0  & 1 & Set\\
			\hline
			1  & 1 & ? & Not allowed  \\
			\hline
		\end{tabular}
		\label{Tab3}
	\end{center}
\end{table}

\begin{figure}[!h]
	\centering
	\includegraphics[width=0.8\linewidth]{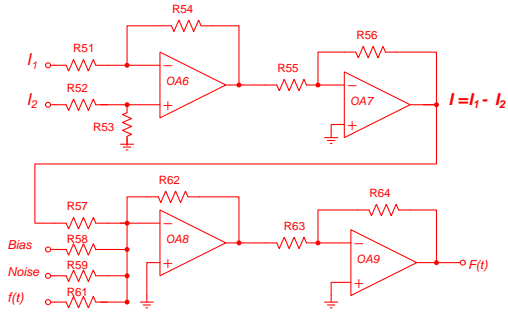}
	\caption{Set-Reset circuit realization for generating the driving signal $ F(t) $. Here we use four op-amps (OA6-OA9) and 14 resistors $R51-R64=10K\Omega$ }
	\label{fig18}
\end{figure}

\begin{figure}[!h]
	\centering
	\includegraphics[width=0.8\linewidth]{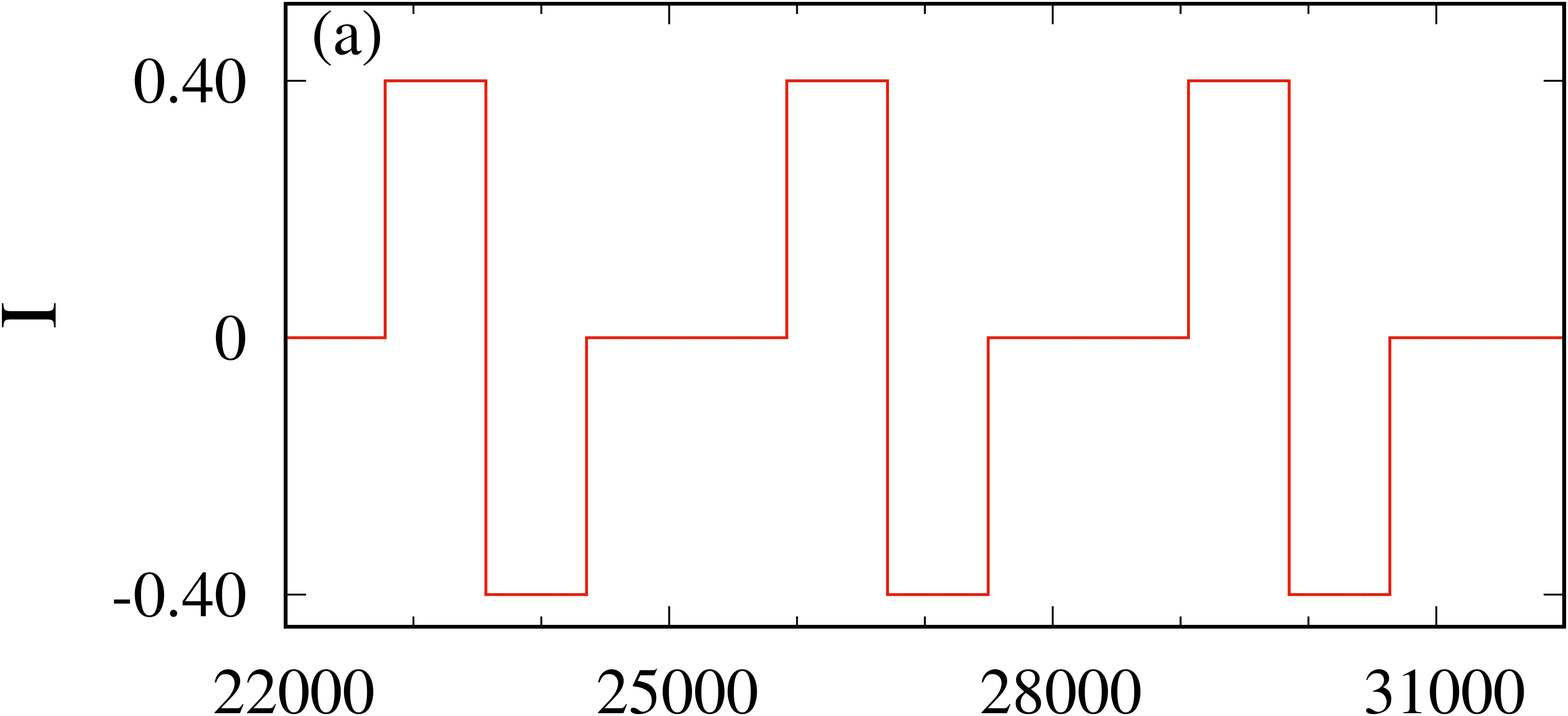}
	\includegraphics[width=0.8\linewidth]{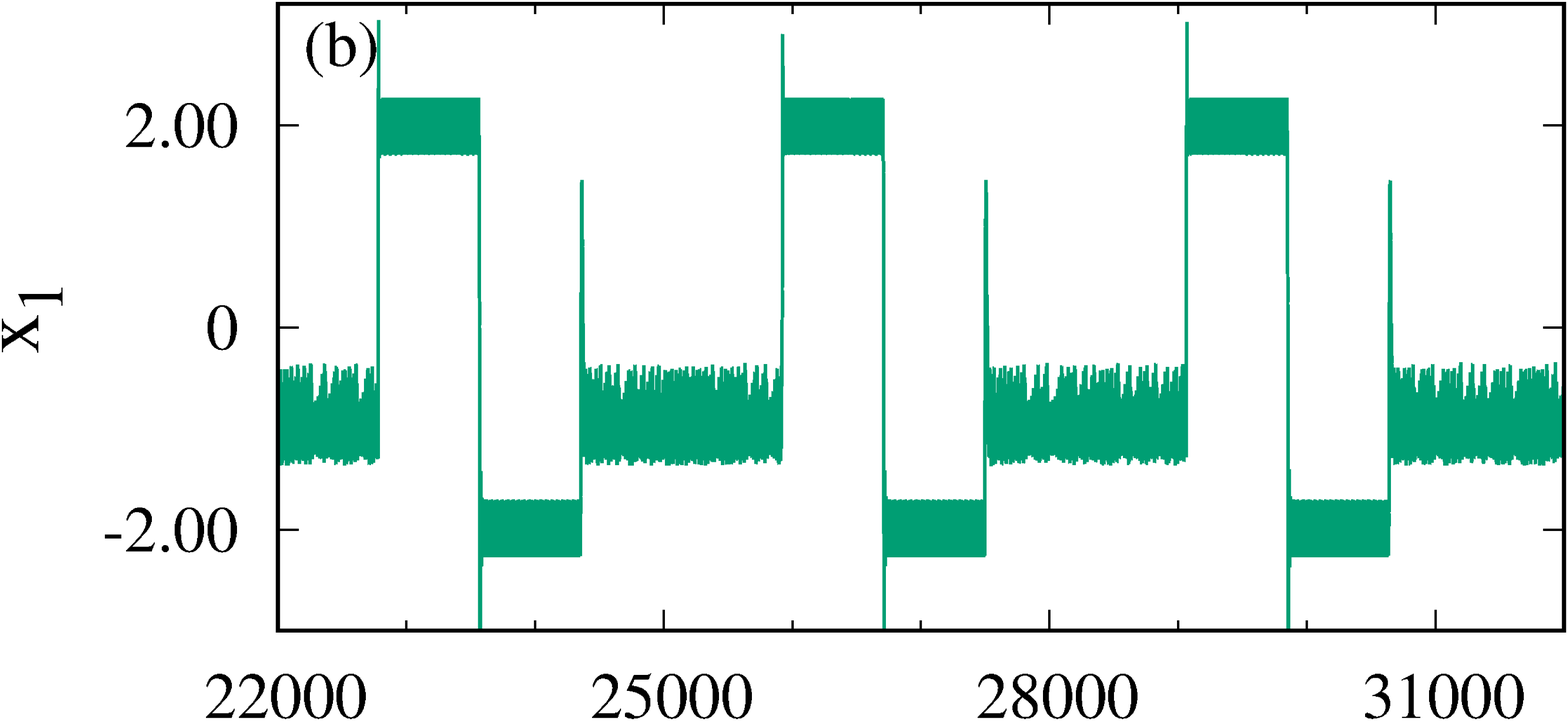}
	\includegraphics[width=0.8\linewidth]{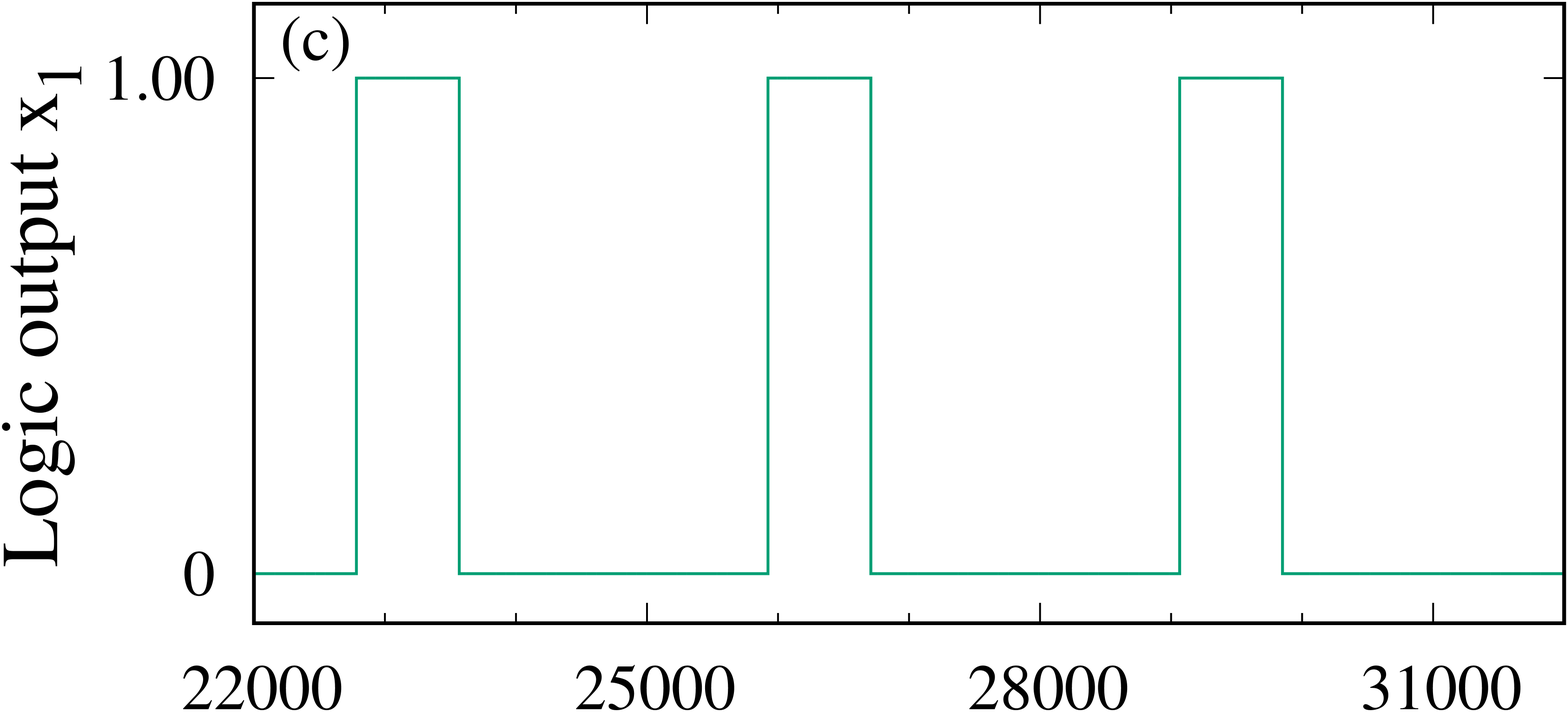}
	\includegraphics[width=0.8\linewidth]{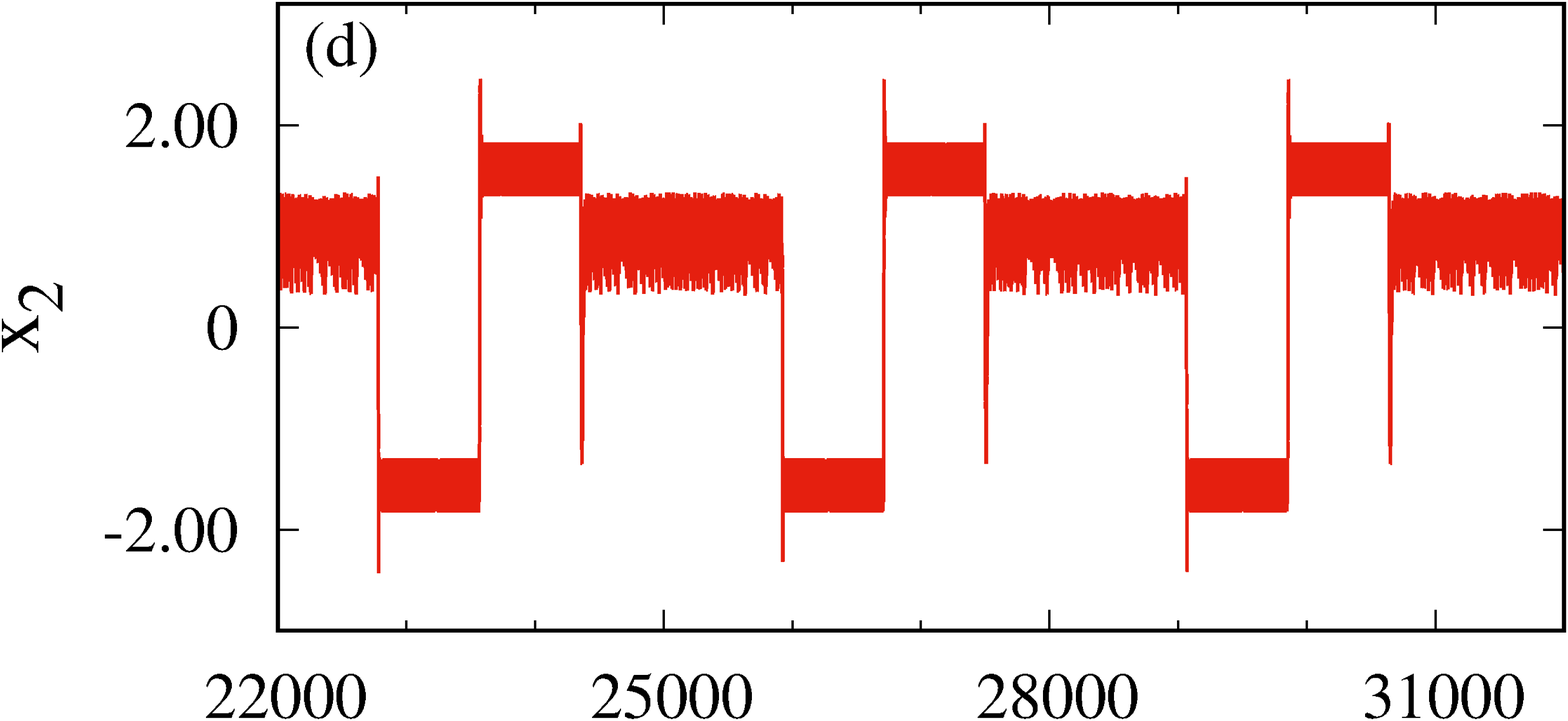}
	\includegraphics[width=0.8\linewidth]{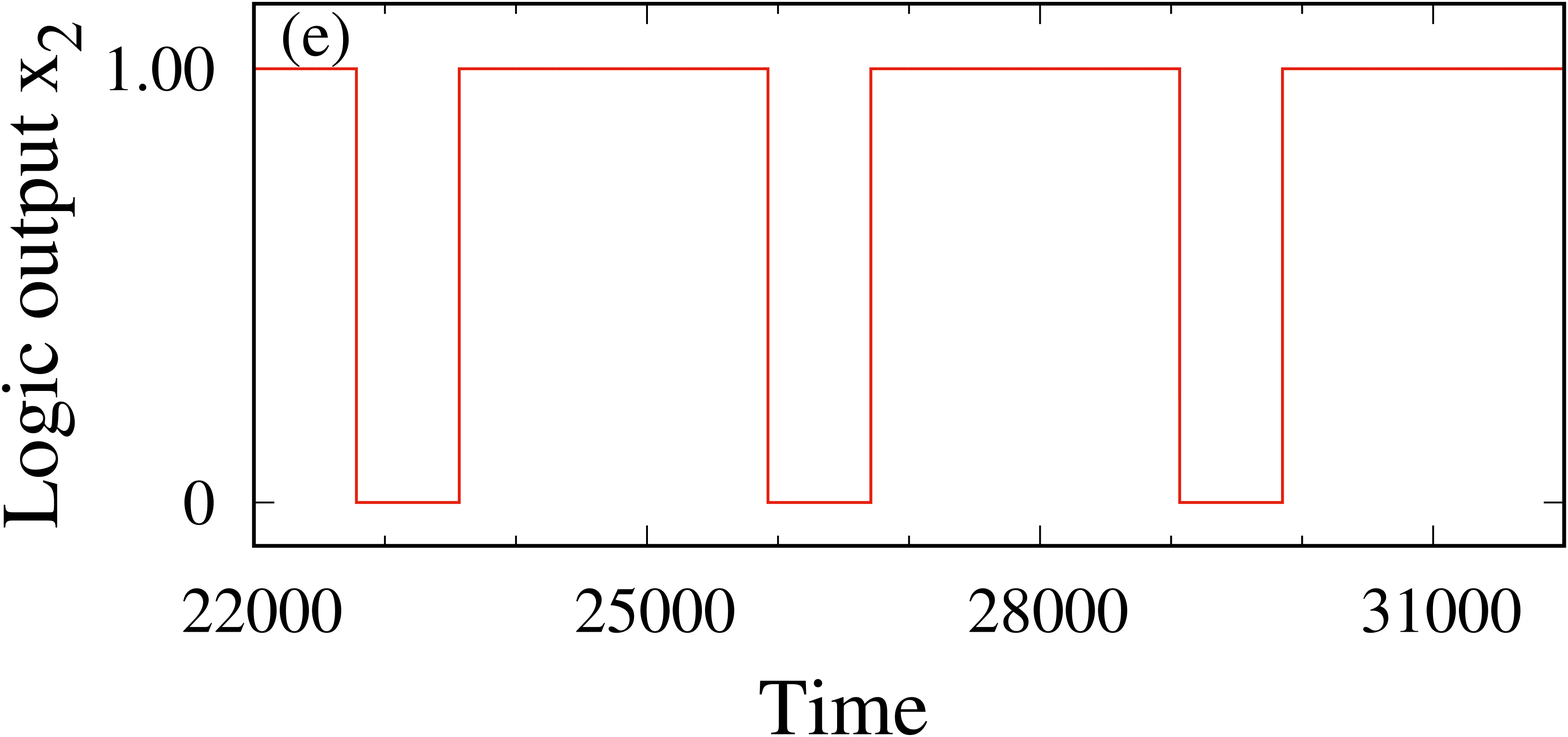}
	\caption{Panel (a) shows a 3-level subtraction of two logic inputs $ I=I_{1}-I_{2} $, $  I=0.1 $ for $ (1,0) $, $ I=0.0 $ for $ (0,0)/(1,1) $ and $ I=-0.1 $ for $ (0,1) $ sets. Panels (b) and (d) are the desired  S-R latch regime of $ x_{1}(t) $ and $ x_{2}(t) $ under periodic forcing $f=0.1$, with fixed parameters $\Delta=0.2$ and $E=0.0$ and \textcolor{black}{panels (c) and (e) represent the corresponding logic outputs of panels (b) and (d). }}
	\label{fig13}
\end{figure}

\begin{figure}[!h]
	\centering
	\includegraphics[width=1.0\linewidth]{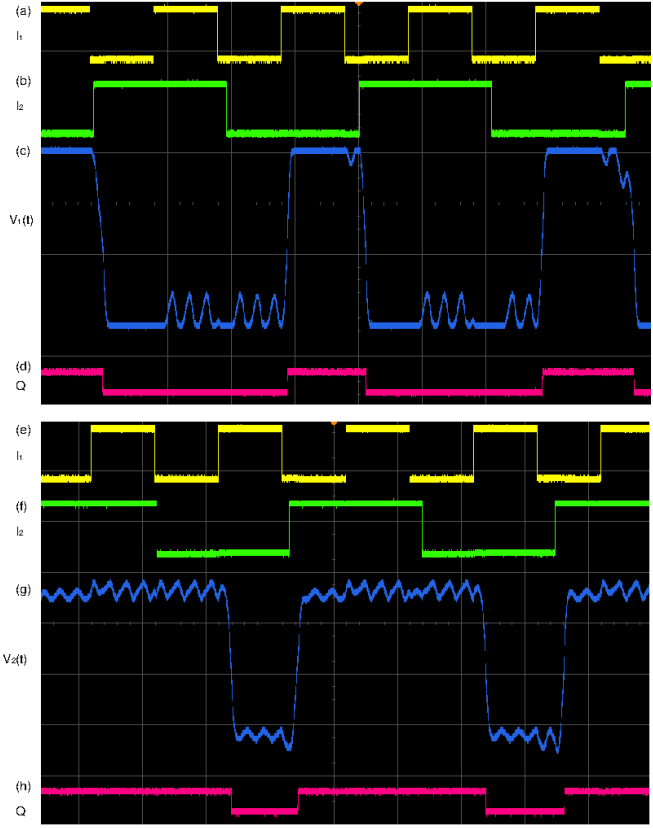}
	\caption{Realization of the Set-Reset memory latch in experimental electronic circuit. Panels (a) \& (e) and (b) \& (f) are the inputs of  $ I_{1} $ and  $ I_{2} $ (if the logic input is 0, it corresponds to $ -100mV $ and $ +100mV $ is considered as the logic input 1). \textcolor{black}{Panels (c) \& (d)  show the Set-Reset memory latch regime and corresponding complementary memory latch output is shown in the panels (g) \& (h).  The experimental output was observed in an Agilent DSO 7014B without bias. }}
	\label{fig14}
\end{figure}

\begin{figure}[!h]
	\centering
	\includegraphics[width=0.9\linewidth]{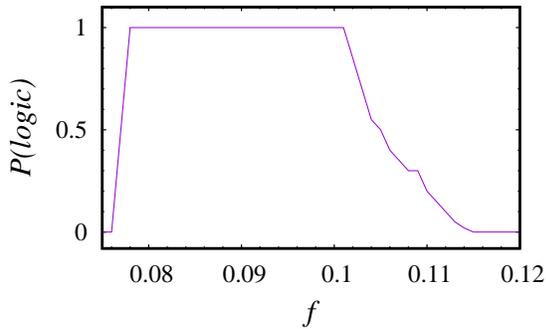}
	\caption{The probability distribution of obtaining logical behavior for different values of $f$ with absence of bias value $ E=0.00 $ for SR memory latch.}
	\label{fig25}
\end{figure}

Apart from designing logic gates many efforts have been made to exploit nonlinear systems to construct memory devices. Chaos based SR flip-flop from two cross coupled Chua's circuit, SR flip-flop using reconfigurable analog block, set-reset latch in noise assisted bistable system, R-S flip-flop from two-cross coupled quasiperiodically driven MLC circuit are few examples for constructing memory latch using nonlinear systems\cite{cafagna2006chaos, campos2012set, venkatesh2017design, sathish2020realisation}. In the present work, we demonstrate how the simple SC-CNN based periodically driven MLC circuit \eqref{equ8} can produce a consistent and complete RS flip-flop. 

In this case, the input streams should be modified. It is quite obvious from the truth table of an SR flip-flop that (0,1) and (1,0) input sets are quite different, since these sets produce different output states (see Table.\ref{Tab3}). As a result we encode the inputs in a different way. Here the first input $ I_{1} $ takes `1' when the logic input is `+1' while it is `0' when the logic input is `0'. But the second input  $ I_{2} $ takes the value `1' when the logic is `0' while `0' for logic `+1'. This could be achieved by applying the NOT operation to $ I_{2} $. For this case also the input streams $(I_1,I_2):(0,0),(0,1),(1,0),$ and $(1,1)$ represent to values of $0,-1,1,$ and $0$ values. Here, $ (1,1) $ set is a restricted one. For example, if $\Delta=0.2$, both the inputs $ I_{1}=I_{2}= $ -0.2 for logical input `$0$' and the values $ 0.2 $ for logical input `$1$'. The input signal $I=I_{1}-I_{2}$ is thus a three-level wave form : $ -0.4 $ for (0,1), $ 0 $ for (0,0)/(1,1)  and $ 0.4 $ for (1,0) inputs sets. 

When we apply above mentioned input streams (Figs.\ref{fig13}(a) and \ref{fig13}(b)) to the system \eqref{equ8}, it is possible to realize the S-R flip-flop in the circuit. As usual, the logical output can be obtained by assuming that the output is to be the value `$ 1 $' for $ x>0 $ and 0 if $ x<0 $. S-R flip-flop operation is clearly shown in Figs.\ref{fig13}(c) \& \ref{fig13}(d) and \ref{fig13}(g) \& \ref{fig13}(h). When $ I_{1} $ and $ I_{2} $ inputs are in the low states, it is found that the system response $ x_{1}(t) $ remains unchanged. When $ I_{1} $ is in the low state and $ I_{2} $ is in the high state, it is observed that the response of the system $ x_{1}(t)<0 $ and hence it is assumed that the logical input is low. This makes the system as a latch with reset condition. When the input $ I_{1} $ is in the high state while $ I_{2} $ is low, it is realized that the output of the system is $ x_{1}(t)>0 $ and thus it is assigned the logical value `$ 1 $'. In a similar way, the output of the dynamical variable $ x_{2}(t) $ (Fig.\ref{fig13}(g)), provides the inverted output of Fig.\ref{fig13}(c) and it behaves as an active high RS flip-flop. Usually in digital circuits, active low and active high RS flip flops are obtained by two cross coupled NAND and NOR gates. But in the present case, these two active high and low RS flip flops are obtained through two dynamical variables of the circuit. Hence the considered circuit is not only providing logical gates but also acts as a sequential circuit to provide RS flip flop operations of various categories such as active low and active high RS flip flops. To realize the Set-Rest latch experimentally, we design the system \eqref{equ8} with two inputs $ I_{1} $ and $ I_{2} $. In the case of logic gates as discussed earlier, two inputs are added by an op-amp summing amplifier, whereas in the case of Set-Reset latch case, the two inputs are subtracted by an op-amp subtracting amplifier. In the circuit of Fig.\ref{fig18}, the signal $ F(t) $ is generated by a set of op-amp summing amplifier by adding the resulting signal $ I(t) $, external bias voltage, external noise signal and sinusoidal signal. Fig.\ref{fig14} verifies the Set-Reset latch behavior in our electronic circuit. \textcolor{black}{ The logical output `$ Q $' in Figs.\ref{fig14}(d)/\ref{fig14}(h) are obtained as $ v_{1}/v_{2} $ by feeding the output response to an appropriate comparator circuit \cite{campos2010simple}.}

\textcolor{black}{Further, we can quantify the process of obtaining a given logic output by calculating $ P(logic) $ as shown in Fig.\ref{fig25} where it is observed the Set-Reset memory latch is realized in an optimal width of forcing amplitude `$ f $'. It is interesting to note that the logic operations which are realized for subthreshold input results in an optimal window of forcing value where $ P(logic)$ tends to $ 1$.}

\begin{figure}[!h]
	\centering
	\includegraphics[width=0.8\linewidth]{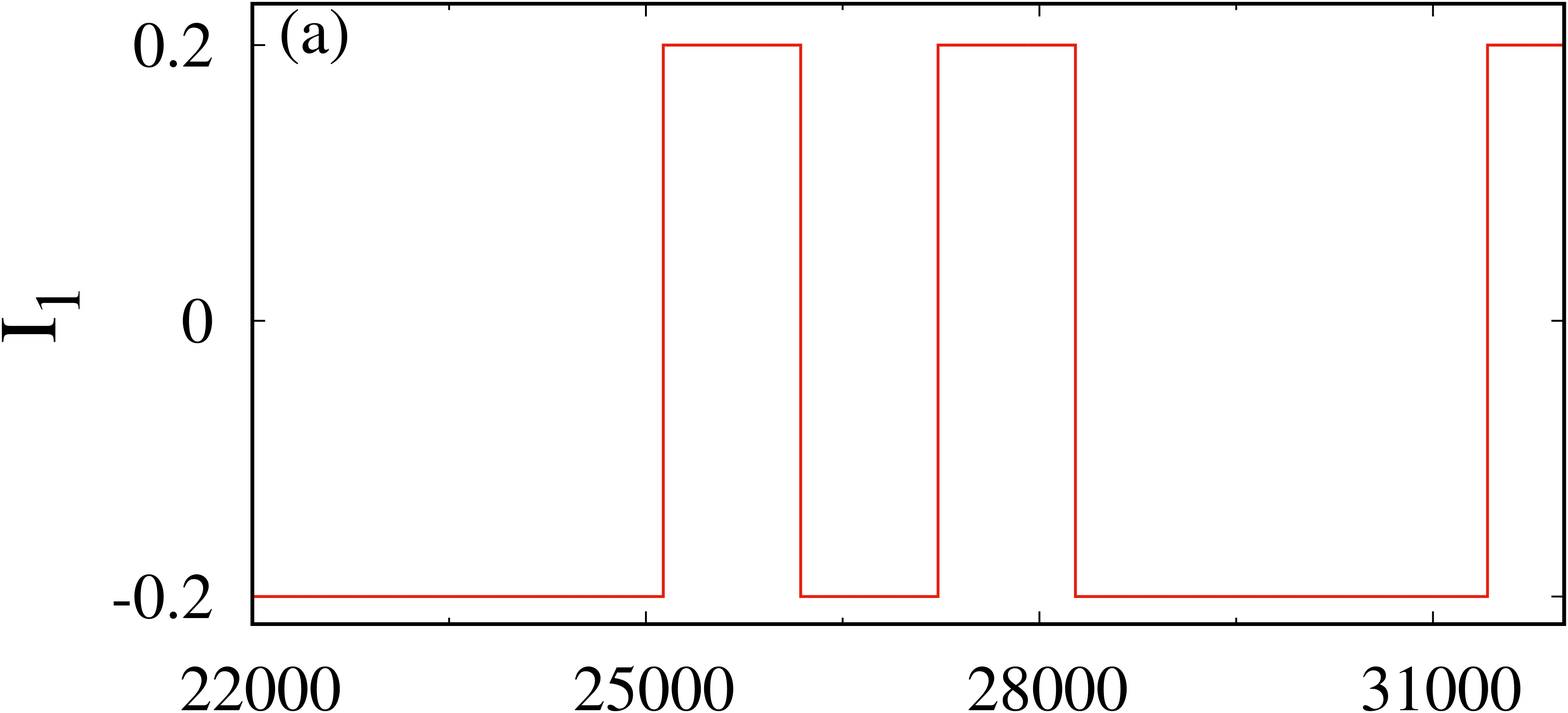}
	\includegraphics[width=0.8\linewidth]{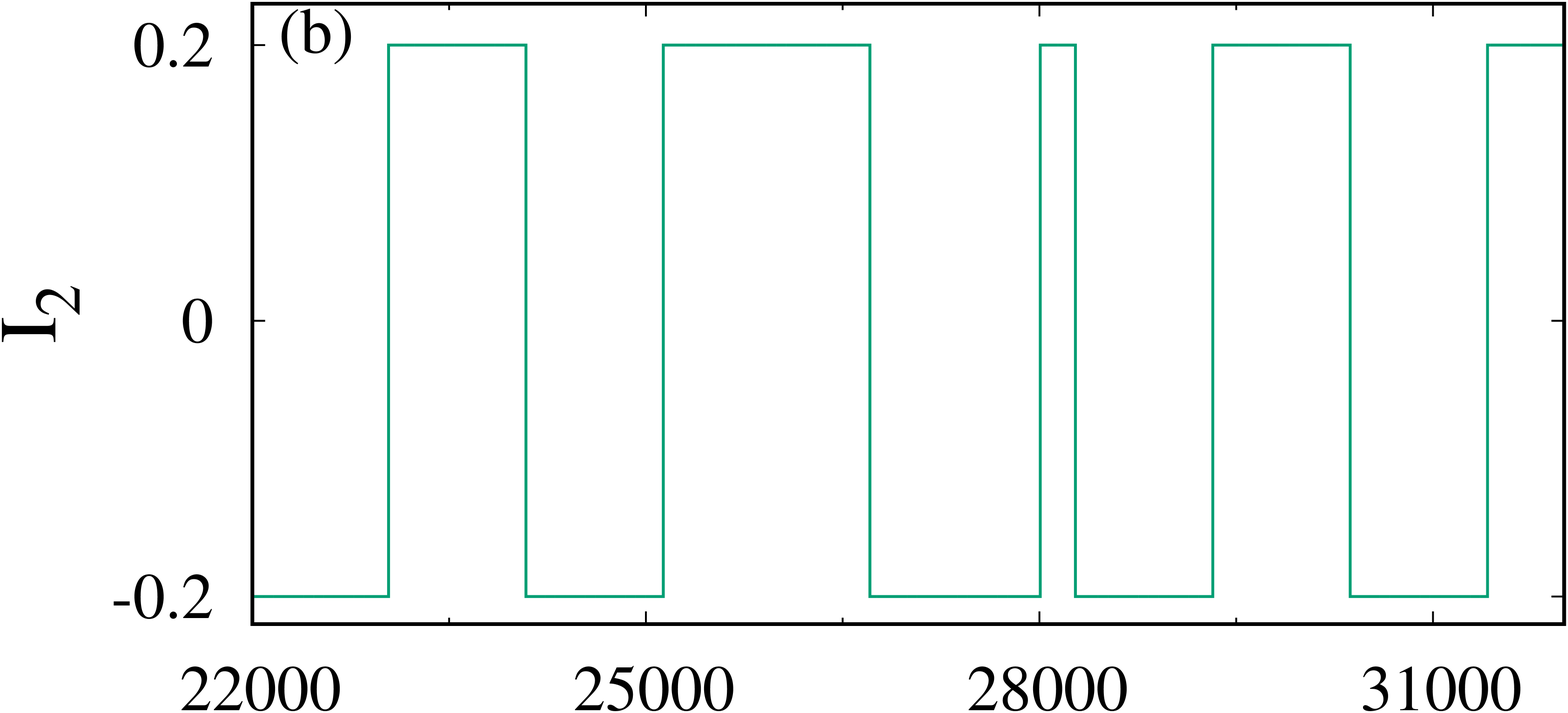}
	\includegraphics[width=0.8\linewidth]{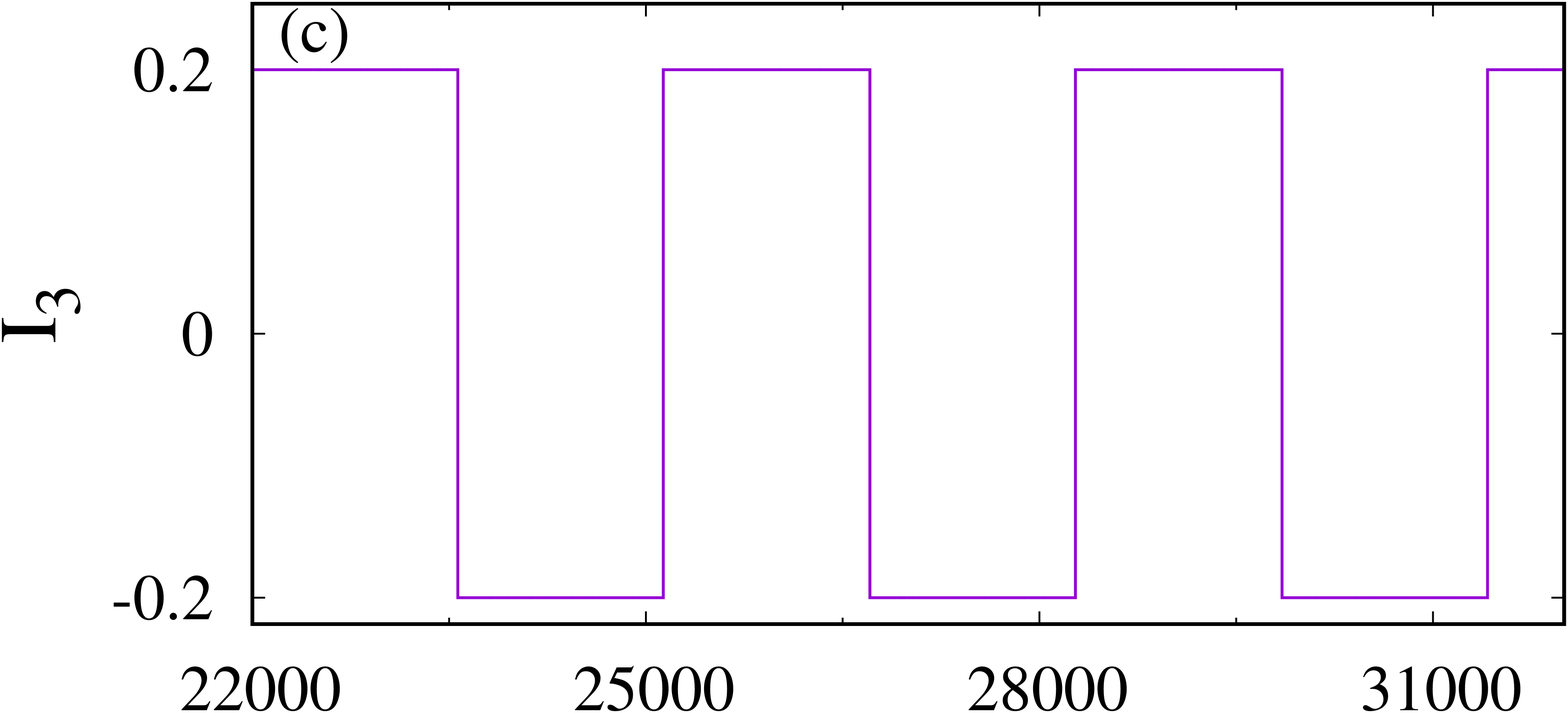}
	\includegraphics[width=0.8\linewidth]{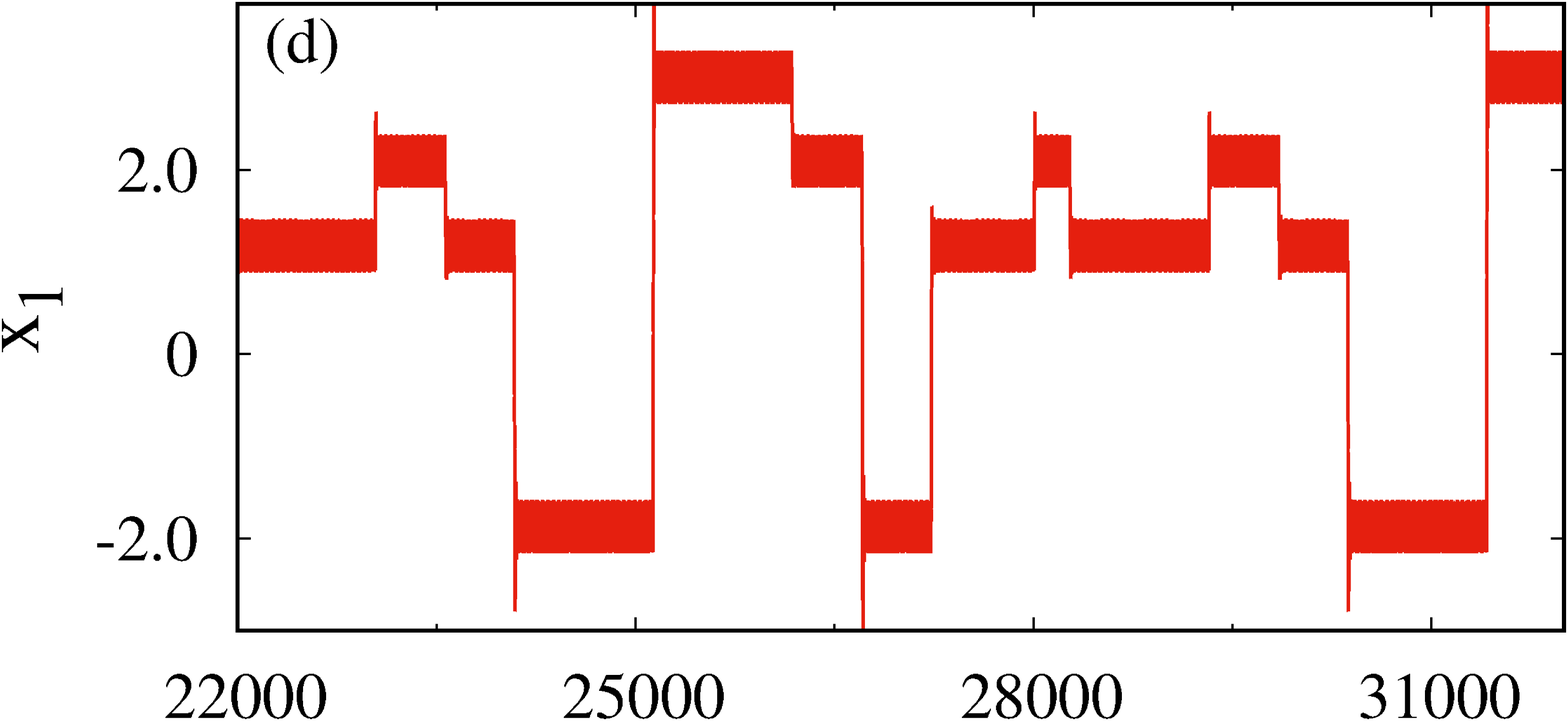}
	\includegraphics[width=0.8\linewidth]{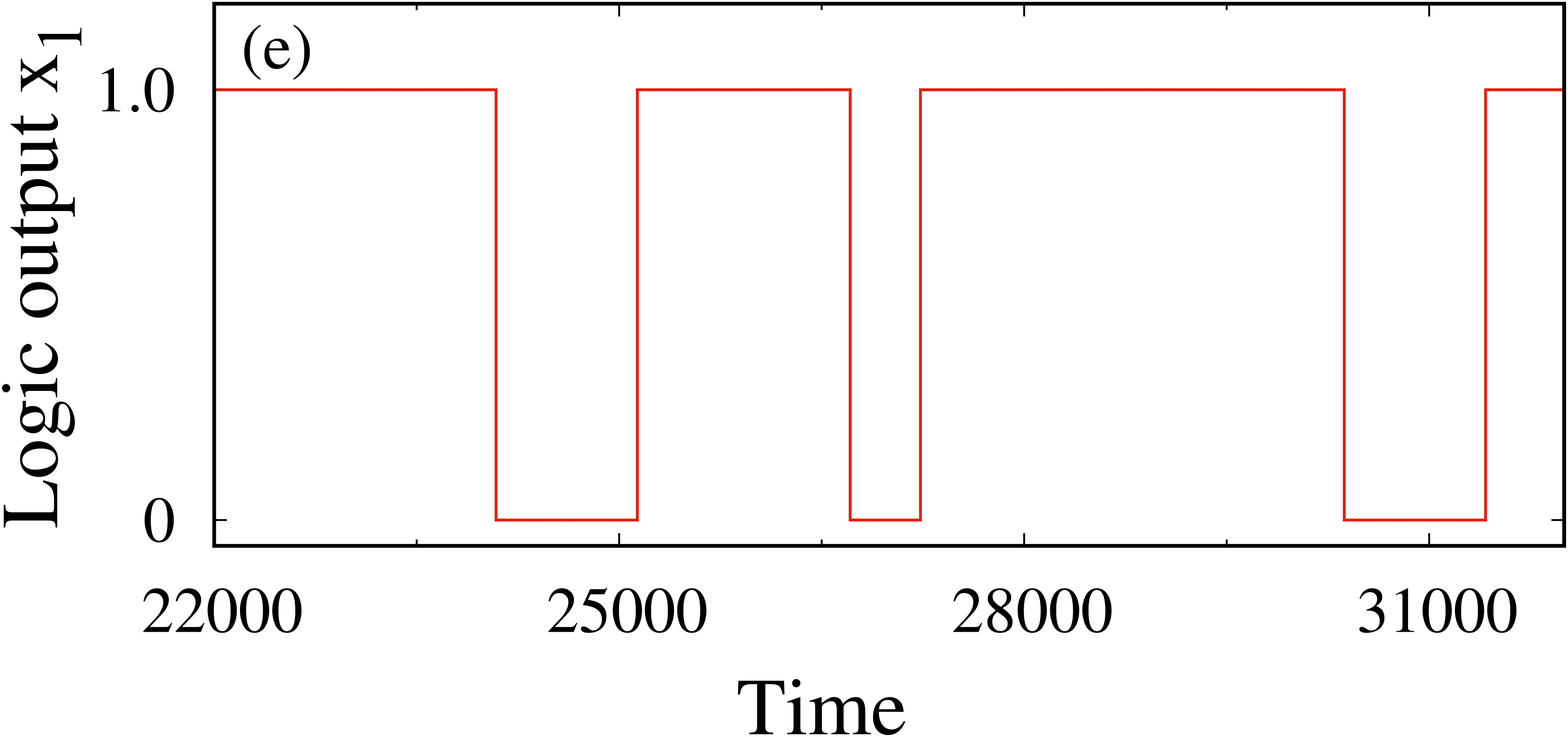}
	\caption{Panels (a)-(c) show the three different logic inputs of $I_{1}$, $I_{2}$ and $I_{3}$, respectively. Input $I_{1}=I_{2}=I_{3}=-0.2$ when the logic input is $'0'$ and $I_{1}=I_{2}=I_{3}=+0.2$ when the logic input is $'1'$. Panel (d) represents the corresponding dynamical response OR logic gate of the system $ x(t) $ under periodic forcing with $f=0.1$ and bias $ E=0.25 $ and \textcolor{black}{ panel (e) represents the corresponding logic output of panel (d).}}
	\label{fig21}
\end{figure}

	\begin{figure}[!h]
		\centering
		\includegraphics[width=1.0\linewidth]{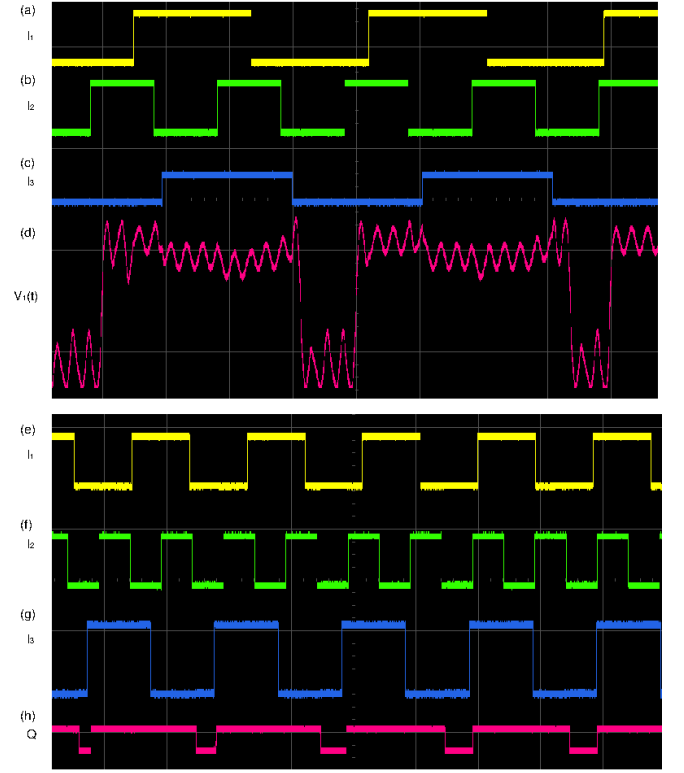}
		\caption{Realization of the OR logic gates in experimental electronic circuit. Panels (a)-(c) and (e)-(g) show the three different logic inputs $ I_{1}, I_{2} \& I_{3}$ (if the logic input is 0, it is  considered as $ -100mV $ and $ +100mV $ is considered as the logic input 1). Panels (d) and (h)  clearly indicate the OR  logic output (when $ x_{1}>0 $ it is considered as logic input 1, and $ x_{1}<0 $ the logic input is considered as 0). \textcolor{black}{The bias voltage is fixed as $ 100 mV $ and the experimental output was observed in an Agilent DSO 7014B.}}
		\label{fig20}
	\end{figure}

\begin{table}
	\begin{center}
		\caption{The truth table for realization of three input OR gate} 
		\vspace{0.2cm}
		\begin{tabular} {|c| c| c| c| c|c|}
			\hline
			\textbf {$ I_{1} ~I_{2}~ I_{3} $} & \textbf{Left well} & \textbf{Center well} & \textbf{Right well} \\
			\hline
			\hline
			0~0~0  & ON  & OFF & OFF \\
			\hline
			0~0~1  & OFF  & ON & ON  \\
			\hline
			0~1~0  & OFF  & ON & ON  \\
			\hline
			0~1~1  & OFF  & ON & ON  \\
			\hline
			1~0~0  & OFF  & ON & ON \\
			\hline
			1~0~1  & OFF  & ON & ON \\
			\hline
			1~1~0  & OFF  & ON & ON \\
			\hline
			1~1~1  & OFF  & ON & ON \\
			\hline
			\hline
			
		\end{tabular}
		\label{Tab4}
	\end{center}
\end{table}

\subsection{Logic responses even in three-input configuration}

Next we consider the question  whether we can extend the scope of this nonlinear circuit system to three inputs or higher inputs?. The answer is yes and we have found that without altering any of the parameters, the system \eqref{equ8} admits logic responses even when feeding three-inputs. In particular, the system produces logic behavior when we feed three inputs $ I_{1} $, $ I_{2} $ and $ I_{3} $.

To be specific, for obtaining the OR gate, we fix the numerical parameters as $I_{1}=I_{2}=I_{3}=0.2$, $E=0.25$ and $f=0.1$ (experimental parameters: $ I_{1} = 100mV $, $ I_{2}=100mV $, $ I_{3}=100mV $, $ f = 2.65KHz $, and $ bias=100mV $). The first three panels of Figs.\ref{fig21} \& \ref{fig20} show the three different inputs $ I=I_{1}+I_{2}+I_{3} $ while the fourth panel corresponds to logical OR dynamics. Hence for states (0,0,0) the output response of the system is assumed to be the logical `0' since the state variable $x_{1}$ oscillates in the $D_{-}$ region and for other states the output response is to be logical '1' value because the state variable $x_{1}$ oscillates in the $D_{+}$ and $ D_{0} $ regions [see Figs.\ref{fig21}(d) \& \ref{fig20}(d)] and also the corresponding logic output represent in Figs.\ref{fig21}(h) \& \ref{fig20}(e). \textcolor{black}{The logical output `$ Q $' in Figs.\ref{fig20}(d)/\ref{fig20}(h) are obtained as $ v_{1}/v_{2} $ by feeding the output response to an appropriate comparator circuit(see Table.\ref{Tab4}).} Switching the bias voltage to $E=-0.25 (bias=-100mV)$, one can obtain three-input AND gate. However, three-input XOR and XNOR are not possible since the circuit is a three-segment piecewise linear system. To realize the three-input XOR and XNOR gates, one has to employ a minimum of four-well potential nonlinear system.

\begin{figure}[!h]
	\centering
	\includegraphics[width=0.9\linewidth]{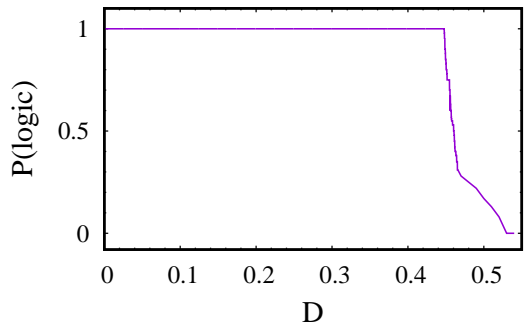}
	\caption{The probability distribution of obtaining logical behavior for different values of $D$ with  bias value $ E=0.01 $ for OR/NOR gates and $ E=-0.01 $ for AND/NAND gates.}
	\label{fig16}
\end{figure}

\subsection{Effect of noise}

Now, we investigate the effect of noise on the logic behavior of the circuit system \eqref{equ8}. In particular, we ascertain the robustness of the logic response with respect to ambient noise. That is, we have to analyze whether the logic response persists even in the presence of noise or not. For this propose, we calculate the probability P(logic) of getting logic gates for different noise strength levels, with strength $ D $. Essentially P(logic) denotes the ratio of the total number of successful runs to the total number of runs. If the system exhibits the desired logic output in response to all the logic inputs, P(logic) is assigned a value `1', otherwise it is treated as `0'. Our numerical simulation, P(logic) for the circuit \eqref{equ8}, is obtained by sampling 1000 runs of the given input set and this process is repeated for 500 such sets. It is clearly indicated in Fig.\ref{fig16}, that for noise strength $ 0 < D < 0.447 $, the system exhibits logic behavior. Beyond $ D>0.447 $, the system loses its logic behavior. Thus, it is clearly demonstrated that the robustness of logical response of the system continues even in the presence of noise originating due to electronic components or any other external factors.

\section{Conclusion}
\label{sec6}

We have investigated the effect of logic square wave signals on the SC-CNN of MLC circuit system. Exploiting the hopping of attractors generated by this circuit, we have found that the circuit can emulate all basic logic operations corresponding to the logic gates OR, AND, NOR, NAND, XOR and XNOR and memory latch element RS flip-flop. It is also shown that all these logic behaviors are tolerant to noise. We have extended the scope of obtaining the logic gates by even feeding higher order inputs, specifically three-inputs. Results obtained by numerical simulations are in good agreement with experimental demonstration. Our study throws some useful light in replacing the existing computer technology with a minimal hardware.

\textcolor{black}{It is clearly evident from our studies that the SC-CNN model of the MLC circuit can function as logic gates as well as memory latch even for small-amplitudes of the input signal. As a result, it consumes only a small power. Further, the same circuit can be used to produce a logic output corresponding to three-input logic. Thus this circuit may be used as two-input and three-input logic gates as well. The low active RS flip-flop and the high active RS flip-flop are constructed by cross-coupling of two NAND gates and cross-coupling of two NOR gates. Hence, the same MLC circuit can produce low active RS flip-flop through one state variable and high active RS flip-flop through another state variable. Further, the circuit can potentially operate in noisy environments.\\
Thus, the circuit can function not only in a noisy atmosphere but can also be capable of utilizing its resources by configuring the circuit into two-input, as well as three input gates and also as a memory device depending on the requirement. For instance, if we need to run a program, the circuit will mimic as logic gates of two inputs or three inputs, and on the other hand, for processing of data, the circuit will mimic as memory device which enables one to utilize it optimally. Apart from the computer architecture, the circuit morphs into a memory latch which may lead to acting as an electronic switching-de bouncer. }

In this paper, we have proved that the underlying chaotic trajectories, which hop in different quadrants of the phase-space, induce parallel logic operations and memory latch in the SC-CNN of the MLC circuit and this feature may be considered for the replacement of currently existing processors and also as a remedy for Moore's law\cite{kia2017nonlinear}. Further one has to study the important features of this circuit such as access clock time, access energies and cell feature size and compare them with the existing technology, provide technically feasibility of the circuit for design and development of dynamic architecture. We hope to pursue these aspects in the near feature.

\section*{AUTHOR’S CONTRIBUTIONS}		
All authors contributed equally.		

\section*{Acknowledgment} 
P.A. and A.V. acknowledge the DST-SERB for providing financial support for a research project under Grant No.EMR/2017/002813. M.S. sincerely thanks the Council of Scientific \& Industrial Research, India for providing a fellowship under SRF Scheme No.08/711(0001)2K19-EMR-I. A.V. also acknowledges the DST-FIST for supporting experimental work under Grant No.SR/FST/College-2018-372(C). M.L. acknowledges the DST-SERB Distinguished Fellowship program under Grant No.SB/DF/04/2017 for financial support.

\section*{DATA AVAILABILITY}
The data that support the findings of this study are available from the corresponding author upon reasonable request.

\end{document}